\documentclass[twocolumn,superscriptaddress,floatfix,preprintnumbers,prl,nofootinbib]{revtex4-2}
\usepackage[colorlinks=true,breaklinks=true]{hyperref}
\usepackage[utf8]{inputenc}
\usepackage{float}
\hypersetup{allcolors=[rgb]{0.0 0.0 0.6},linkcolor=[rgb]{0.75 0.05 0.05}}
\usepackage{amsmath,amssymb}
\usepackage{mathtools}
\usepackage[normalem]{ulem}
\usepackage{bm}
\usepackage{physics}
\usepackage{siunitx}
\DeclareSIUnit \parsec {pc}
\DeclareSIUnit \years {yr}
\DeclareSIUnit \electronvolt {eV}
\usepackage[dvipsnames]{xcolor}
\usepackage{float}
\usepackage{graphicx}
\usepackage{epsfig}
\usepackage{letltxmacro}
\LetLtxMacro{\oldcite}{\cite}
\renewcommand{\cite}[1]{\mbox{\oldcite{#1}}}
\usepackage{orcidlink}
\usepackage{fontawesome5}
\usepackage{xspace}

\makeatletter
\def\@hangfrom@section#1#2#3{\@hangfrom{#1#2}{#3}}
\def\@hangfroms@section#1#2{#1#2}
\makeatother

\newcommand{\mn}{\ensuremath{m_\bullet}}   
\newcommand{\Mb}{\ensuremath{M_\bullet}}   

\usepackage{longtable,booktabs} 

\usepackage{orcidlink}

\long\def\exclude#1{}

\DeclareSymbolFont{starfontsym}{OT1}{sts}{m}{n}
\DeclareMathSymbol{\mathTerra}{\mathord}{starfontsym}{76}

\newcommand{\beq}{\begin{equation}}
\newcommand{\eeq}{\end{equation}}

\def\ga{\,\,\raise0.14em\hbox{$>$}\kern-0.76em\lower0.28em\hbox
{$\sim$}\,\,}

\long\def\exclude#1{}

\allowdisplaybreaks

\bibpunct{[}{]}{,}{n}{}{,}

\hypersetup{
    colorlinks=true,
    citecolor=[rgb]{.1, .7, .6},
    linkcolor=[rgb]{.2, .55, .95},
    filecolor=magenta,
    urlcolor=[rgb]{.1, .7, .6},
}

\usepackage{appendix}

\newcommand{\githubmaster}{\href{https://github.com/andrea0292/s2-to-lisa/}{\faGithub}\xspace}

\makeatletter
\renewcommand\paragraph{\@startsection{paragraph}{4}{\z@}%
  {3.25ex \@plus1ex \@minus.2ex}%
  {-1em}%
  {\normalfont\normalsize\bfseries}}
\makeatother

\begin{document}

\title{From S2 to LISA: Astrometric Bounds on Extreme-Mass-Ratio Inspirals and Bursts}

\author{Andrea Caputo \orcidlink{0000-0003-1122-6606}} \email{andrea.caputo@cern.ch}
\affiliation{Department of Theoretical Physics, CERN, Esplanade des Particules 1, P.O. Box 1211, Geneva 23, Switzerland}
\affiliation{Dipartimento di Fisica, ``Sapienza'' Universit\`a di Roma \& Sezione INFN Roma1, Piazzale Aldo Moro
5, 00185, Roma, Italy}
\affiliation{Department of Particle Physics and Astrophysics, Weizmann Institute of Science, Rehovot 7610001, Israel}

\author{Valerie Domcke \orcidlink{0000-0002-7208-4464}} \email{valerie.domcke@cern.ch}
\affiliation{Department of Theoretical Physics, CERN, Esplanade des Particules 1, P.O. Box 1211, Geneva 23, Switzerland}

\begin{abstract}
Stellar orbits around the massive black hole at the center of our galaxy provide a unique local probe of the compact-object population in the Galactic Centre and, consequently, of the sources of millihertz gravitational waves: periapse passages lead to extreme-mass-ratio bursts (EMRBs) while successful captures lead to extreme-mass ration inspirals (EMRIs). In this paper we use recent astrometric limits from the GRAVITY observatory on perturbations of the orbit of the star S2 to place upper limits on the normalisation of a stellar-mass black-hole cusp within ${\sim}0.02\,\mathrm{pc}$. For a benchmark $10\,M_\odot$ Bahcall--Wolf population anchored to this data, we obtain upper limits of ${\sim}2.4\times10^{2}\,\mathrm{Gyr}^{-1}$ on the EMRI rate and ${\sim}0.2\,\mathrm{yr}^{-1}$ on the detectable EMRB rate in the Milky Way, which fall within the broad range of previous theoretical estimates. Assuming a self-similar scaling of the cusp normalisation with central black-hole mass, we extend this calibration to cosmological populations. The resulting EMRI background is detectable by LISA across all scenarios considered, whereas the flatter EMRB background can reach LISA sensitivity when mass segregation is efficient in low-mass galactic nuclei. Our results highlight the complementarity of precision stellar astrometry and millihertz gravitational-wave observations. \githubmaster

\end{abstract}

\preprint{CERN-TH-2026-176}
\maketitle

\section{Introduction.} Extreme--mass--ratio inspirals (EMRIs), the gradual capture of a compact object by a massive black hole (MBH), are among the flagship targets of the Laser Inteferometer Space Antenna LISA~\cite{Gair:2017ynp,LISA:2024hlh, Babak:2017tow, Amaro-Seoane:2007osp, Bonetti:2020jku}. Their long-lived, information-rich waveforms will enable precision measurements of strong-field gravity and the demographics of MBHs surrounded by compact remnants, in particular stellar-mass black holes (sBHs).
The same environments source extreme--mass--ratio bursts (EMRBs), short
gravitational-wave (GW) transients emitted during individual periapsis passages of compact objects on highly eccentric orbits~\cite{Berry:2010gt,Berry:2012im, Rubbo2006ApJ, Hopman2007, Oliver:2025irg}. Their higher event rate may put a galactic EMRB event within the reach of the LISA mision~\cite{Rubbo2006ApJ,Hopman2007,Berry:2013poa}, providing a gravitational probe of the inner region of the Milky Way~\cite{Berry:2013ara}.

However, the prediction for both EMRI and ERMB event rates are notoriously plagued by large astrophysical uncertainties~\cite{Babak:2017tow}. The event rate in a given galaxy depends on the poorly constrained phase-space distribution of stellar-mass black holes and other remnants within the inner 
$\lesssim 10$--$100\,\mathrm{mpc}$
of galactic nuclei, on the efficiency of relaxation processes that refill the loss cone, and on the branching between prompt plunges and long-lived inspirals~\cite{Hopman:2005vr, BarOr2016, Amaro-Seoane:2012lgq, Pan:2021ksp, Rom:2024nso, Kaur:2024ofj, Alexander:2008tq}. The extragalactic rate additionally depends on the MBH population model. As a result, published rate forecasts span orders of magnitude even for Milky-Way-like hosts, reflecting uncertainties in cusp formation~\cite{1977ApJ...216..883B} and re-growth, mass segregation~\cite{Alexander:2008tq}, nuclear star-cluster properties, and MBH occupation at low masses~\cite{Greene_2020, Burke:2024wcf}.

A key limitation is that, outside the Milky Way, the relevant scales are fundamentally inaccessible to direct dynamical probes. The EMRI and EMRB-producing region corresponds to milliparsec-to-centiparsec radii, where the potential is dominated by the MBH and where the supply of compact objects is set by relaxation over long timescales. For galaxies other than our own, stellar proper motions and accelerations at these radii cannot be resolved with current instrumentation, and most models therefore rely on extrapolations of idealized cusp solutions or on population-synthesis assumptions calibrated indirectly.

The Galactic Centre (GC) is unique in this respect. It is the only environment where individual stellar orbits can be tracked with sufficient precision to constrain the gravitational potential at $\mathcal{O}(10)\,\mathrm{mpc}$ scales~\cite{Schodel:2002py, Ghez:2008ms, Gillessen_2017, Genzel2010}, covering the radii most relevant to EMRI and EMRBs formation. This landscape is currently being transformed by high-precision astrometry~\cite{GRAVITY2017, GRAVITY:2018ofz, GRAVITY2020, GRAVITY2024}. The GRAVITY and GRAVITY+ instruments have reached the sensitivity required to detect "granularity" in the gravitational potential of the MBH Sgr~A* in the center of our galaxy by searching for subtle orbital residuals -- caused by flybys of individual stellar-mass BHs -- in the highly eccentric orbit of the star S2 orbiting Sgr~A* with a pericenter (closest passage) of only 
0.6~mpc (120~AU)~\cite{Bordoni:2025mli}. Such measurements can provide the first direct, empirical normalization of the local BH population.

In this work, we bridge the gap between local astrometry and cosmological gravitational-wave science. Rather than relying on top-down population synthesis models, we use the potential GRAVITY+ discovery of a sparse, heavy BH cluster to calibrate the EMRI and EMRB event rates. This is particularly relevant for EMRBs, for which detectable \textit{Galactic}
events are indeed plausible. Furthermore, while the Milky Way is not necessarily representative of all galactic nuclei, we extend our results to other MBHs using simple but physically motivated arguments, based on self-similarity and calibrated against numerical simulations. In doing so, we confirm that the EMRI rates reported in the literature are broadly consistent with the current (non-)~detection by GRAVITY, {but that upcoming astrometric data will probe the commonly adopted assumptions on the sBH distribution. Moreover, we show that the stochastic gravitational wave background~(SGWB) from extragalactic EMRBs, typically neglected in astrophysical gravitational-wave projections, may in fact not be negligible, at least under optimistic assumptions. This exercise illustrates how strong local constraints on the remnant population at $\lesssim 0.02\,\mathrm{pc}$ scales can inform expectations for LISA, and motivates joint analyses connecting Galactic-centre astrometry to space-based
gravitational-wave observations.

The remainder of this paper is organized as follows. In Section~\ref{sec:gravity} we translate the limits by GRAVITY on the precession and inclination of S2 to upper bounds on the sBH number density. These results are extrapolated to extragalactic sources based on the scaling arguments presented in Sec.~\ref{sec:scaling}. In Section~\ref{sec:general_formalism} we review the formalism to compute event rates and SGWBs, both for EMRBs and EMRIs. Section~\ref{sec:results} combines all the above to derive our predictions for event rates and SGWBs of both source classes, before concluding in Sec.~\ref{sec:conclusions}. Technical details are collected in several appendices.

\begin{figure*}[t]
  \centering
\includegraphics[width=2\columnwidth]{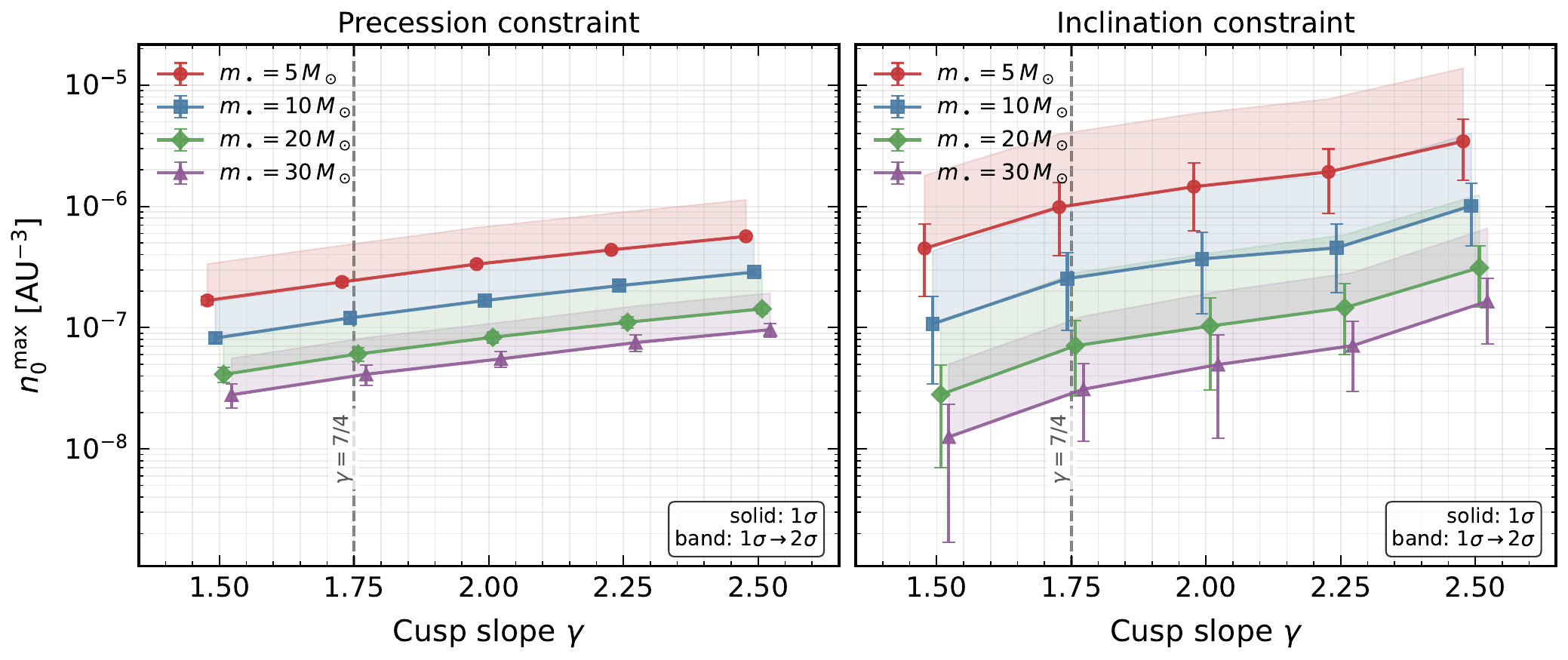}
\caption{Upper limits on the density normalisation $n_0$ of a stellar-mass BH cusp around
Sgr\,A$^*$, derived from GRAVITY observations of S2's orbit. \textit{Left:} constraint from the Schwarzschild precession measurement, with a $1\sigma$
tolerance $\sigma_{\Delta\omega} = 1.35\,$arcmin per orbit set by the precession factor $f_\mathrm{SP} = 1.135 \pm 0.110$~\cite{GRAVITY2024} (i.e.\
$\sigma(f_\mathrm{SP})$ times the relativistic precession $\simeq 12.3\,$arcmin per orbit). \textit{Right:} constraint from the stability of the orbital plane, with $\sigma_{\Delta i} = 40\,$arcsec at $1\sigma$, defined as the inclination
change that displaces S2 on the sky by GRAVITY's astrometric accuracy of $\approx 30\,\mu$as~\cite{GRAVITY2017} at apocentre, after the geometric projection $\sin i\,|\sin(\nu+\omega)|\approx 0.66$.
Each curve is a different perturber mass $m$ (red: $5\,M_\odot$, blue:
$10\,M_\odot$, green: $20\,M_\odot$, purple: $30\,M_\odot$). Solid lines with
error bars show the $1\sigma$ upper limit on $n_0$ at each cusp slope $\gamma$; shaded bands extend to the $2\sigma$ limit. Error bars propagate the uncertainty
on the fitted scaling coefficients $A$ and $B$ through the $N_\mathrm{max}$ derivation. For each $(\gamma,\mn)$ the simulation sweeps over $N_p$ with $40$ Monte Carlo realisations per point, fitting $|\Delta\omega| = A\,N_p$ and $\Delta i = B\,\sqrt{N_p}$.}
\label{fig:n0_s2}
\end{figure*}

\section{Calibrating sBH profiles with GRAVITY}
\label{sec:gravity}

The astrometric residuals discussed above turn S2 into a dynamical probe for the unseen remnant population, and provide an empirical normalisation that all our subsequent rate estimates rest on. Once a radial profile and a mass spectrum are assumed, the demographics of the cusp reduce to a single number, the normalisation $n_0$ of the density profile, which fixes how numerous the stellar-mass BH population is. This in turn determines the overall amplitude of both the EMRI and EMRB signals (we will see that the rates scale as $\Gamma\propto n_0^\alpha$ with $1<\alpha <2$). Here we make this calibration quantitative, deriving upper limits for $n_0$ from an N-body integration of S2's orbit.

A cusp of stellar-mass BHs imprints itself on the orbit of S2 in two complementary ways. Its \emph{smooth} enclosed mass advances the apsidal precession beyond the relativistic value, while the \emph{granular}, discrete nature of the individual
perturbers (the same flyby-induced residuals that GRAVITY+ is sensitive to~\cite{Bordoni:2025mli}) torques and reorients the orbital plane. The two
channels carry genuinely different information, and we constrain them separately.

Our integrator evolves S2 under (i) the Newtonian potential of Sgr~A$^*$, (ii) the 1PN Schwarzschild precession, and (iii) the gravitational perturbations from $N_p$ stellar-mass BHs on Keplerian orbits. The perturber semi-major axes $a$ are sampled from the power-law profile $n(a)\propto a^{-\gamma}$ between $0.5\,r_{\rm peri}$ and $2\,r_{\rm apo}$ of S2 (i.e.\ $3\cdot10^{-4}$--$2\cdot10^{-2}\,$pc, covering the $\lesssim 0.02\,$pc scales relevant to EMRI formation), with a thermal eccentricity distribution $f(e)=2e$ and isotropic orientations. Each realisation is integrated over two S2 orbital periods ($\simeq 33\,$yr), and we record the additional apsidal precession $\Delta\omega$ and the change in
orbital inclination $\Delta i$ between successive pericentre passages. Running $40$ Monte Carlo realisations per value of $N_p$, we recover the two expected scalings,
\begin{equation}
  |\Delta\omega| = A\,N_p, \qquad \Delta i_{\rm rms} = B\,\sqrt{N_p},
\end{equation}
with $A$ and $B$ both proportional to the perturber mass $m_\bullet$.\footnote{A priori, these perturbers could be both black holes and stars. Since mass segregation~\cite{Alexander:2008tq,Preto:2009kd} leads to a concentration of sBHs in the inner region of the galaxy probed by GRAVITY, we will largely focus on sBH perturbers in the following, if not stated explicitly otherwise.}
The linear growth of $\Delta\omega$ reflects the \emph{coherent} addition of apsidal kicks (precession tracking the smooth enclosed mass $\propto N_p\,m_\bullet$), whereas the $\sqrt{N_p}$ growth of $\Delta i$ reflects a \emph{random walk} of
randomly oriented plane changes, sensitive to the granularity of the discrete population. Requiring the perturbations that S2 acquires to remain within the precision of the GRAVITY measurements fixes, for each channel, the maximum allowed number of perturbers $N_{\max}$, which we convert into an upper limit on $n_0$. For the apsidal advance we adopt a $1\sigma$ tolerance $\sigma_{\Delta\omega}\simeq 1.35\,$arcmin per orbit, the width of the measured Schwarzschild--precession factor $f_{\rm SP}=1.135\pm0.110$~\citep{GRAVITY2024} times the relativistic precession $\Delta\omega_{\rm GR}\simeq 12.3\,$arcmin per orbit. For the reorientation of the orbital plane we set $\sigma_{\Delta i}\simeq 40\,$arcsec, defined as the plane tilt that displaces S2 on the sky by GRAVITY's astrometric accuracy of $\approx 30\,\mu$as~\citep{GRAVITY2017, Bordoni:2025mli}.\footnote{A tilt $\Delta i$ moves the star by
$\sin i\,\,r\,|\sin(\nu+\omega)|\,\Delta i$, where $\nu$ is the true anomaly, which is largest at apocentre; the geometric projection ($\sin i$ together with the orbital-phase factor) reduces
the naive lever arm by $\approx0.66$, so that a $30\,\mu$as displacement corresponds to $\Delta i\simeq 40\,$arcsec.} Inverting $|\Delta\omega|=A\,N_p$ and $\Delta i_{\rm rms}=B\,\sqrt{N_p}$ at these thresholds yields $N_{\max}$ for the two channels.

It is instructive to recast $N_{\max}$ as the mass enclosed between S2's periapse and apoapse, $M_{\mathrm{peri}\to\mathrm{apo}}$, which compares directly with GRAVITY's limit on a smooth extended-mass component within the orbit, $\simeq 4\times10^{3}\,M_\odot$ at $1\sigma$~\citep{GRAVITY2020}, recently
tightened to $\lesssim 1.2\times10^{3}\,M_\odot$~\citep{GRAVITY2024}. From the precession statistic we infer $M_{\mathrm{peri}\to\mathrm{apo}}\lesssim
(1.3\text{--}1.7)\times10^{3}\,M_\odot$ ($1\sigma$) across the entire $(\gamma, \mn )$ grid, essentially independent of the individual perturber mass. Precession responds only to the \emph{smooth} enclosed mass---the very quantity
GRAVITY constrains---so this channel does not add information independent of GRAVITY's mass limit but provides a consistency check, which our bound (built on the same $f_{\rm SP}$ measurement) passes by construction. We verified the agreement explicitly: replacing the discrete cusp by a smooth Plummer component with GRAVITY's own scale ($a=0.3''$) and their $2020$ precision returns $\approx 3.5\times10^{3}\,M_\odot$ within the orbit, in agreement with their $\simeq 4\times10^{3}\,M_\odot$, and the integrator recovers the relativistic
precession to better than $1\%$ ($12.23$ vs.\ $12.26\,$arcmin per orbit). The slightly weaker limit on the central mass (compared to \citep{GRAVITY2024}) reflects the improved $2024$ measurement and the fact that a constraint from S2's precession alone is necessarily looser than GRAVITY's full multi-star orbital fit.

The inclination channel instead carries the \emph{granular} information (the
discrete, randomly oriented plane kicks to which smooth-mass fits are blind) and
is strongly mass dependent. At $1\sigma$ it is tightest for the heaviest perturbers: for $\mn=30\,M_\odot$ it gives $M_{\mathrm{peri}\to\mathrm{apo}}\lesssim
(0.8\text{--}1.4)\times10^{3}\,M_\odot$ over most of the grid, comparable to GRAVITY's direct limit, because fewer heavy objects deliver larger, more easily detected plane changes. For light perturbers the kicks are small and the bound relaxes to $\gtrsim 4\times10^{3}\,M_\odot$ ($\mn=5\,M_\odot$), no longer
competitive with the precession bound. Comparing the two channels, the precession limit is the binding constraint over most of the $(\gamma,\mn)$ grid, while the granular inclination constraint overtakes it only for the heaviest remnants ($\mn\gtrsim 20\text{--}30\,M_\odot$), where the per-object kicks are largest. More details are given in App.~\ref{app:simulations} and~\ref{app:n0_method}.

These limits are collected in Fig.~\ref{fig:n0_s2} as upper limits on $n_0$ versus the slope $\gamma$, for four representative perturber masses. The precession bound (left) depends only weakly on mass ($n_0^{\max}\propto m_\bullet^{-1}$, since it
fixes the mass-independent enclosed mass), whereas the inclination bound (right) separates more strongly ($n_0^{\max}\propto m_\bullet^{-2}$, as $\Delta i=B\sqrt{N_p}$ with $B\propto m_\bullet$), giving the tightest $n_0$ for the heaviest black holes. Solid lines give the $1\sigma$ limits, with error bars
propagated from the fitted coefficients $A$ and $B$; shaded bands extend to the $2\sigma$ thresholds.

Because both coefficients scale as $A,B\propto m_\bullet$, these limits rescale to any assumed remnant mass without re-running the integration: the precession limit scales as $n_0^{\max}\propto m_\bullet^{-1}$ and the granular bound as $n_0^{\max}\propto m_\bullet^{-2}$, so for the rate forecasts we adopt, at each $(\gamma,m_\bullet)$, the tighter of the two---precession for light remnants and
the granular bound for the heavy remnants ($m_\bullet\gtrsim20\text{--}30\,M_\odot$). Naturally, other massive black holes inhabit different environments and may host stellar-mass black hole populations with different normalisations and slopes; the Milky Way,
however, as the only nucleus where direct astrometric monitoring of an individual stellar orbit pins down the cusp, provides a well-motivated anchor for the predictions that follow. 

The vertical line in Fig.~\ref{fig:n0_s2} indicates the  Bahcall Wolf (BW) profile~\cite{1977ApJ...216..883B}, $\gamma = 7/4$, which will serve as a benchmark case for the rest of this paper. This is consistent with the two-species steady-state structure of Ref.~\cite{Rom:2024nso}, in which the sBH density breaks from a steep, star-dominated branch ($\gamma=4$) at large radii into the shallower, sBH-dominated BW branch inside a transition radius
$r_{\rm I}=f_\bullet^{4/5}(m_\bullet/m_\star)^{6/5}r_h$, where $r_h = GM_\bullet/\sigma^2$, is the radius of influence of the MBH (here $\sigma$ is the velocity dispersion), and $f_\bullet$ is the sBH number fraction within the sphere of influence. The BW branch governs the EMRI- and EMRB-producing radii as long as the sBH number fraction exceeds $f_\bullet^{c}\simeq4.5\times10^{-4}\left(\frac{m_\bullet/m_\star}{10}\right)$,
equivalently as long as the relaxation radius falls inside $r_{\rm I}$. Saturating the GRAVITY limit fixes this abundance directly from the data, and integrating the BW cusp to
the S2 bound gives $f_\bullet\simeq1.7\times10^{-2}$,
a factor $\sim40$ above $f_\bullet^{c}$. The MW nucleus is therefore self-consistently sBH-dominated at every observable radius, with the BW branch reaching out to $r_{\rm I}\simeq2.4\times10^{5} \rm AU$.

We close this section by commenting on the assumption of a single perturber mass. In general, one would expect to have a mass spectrum $dN/dm_\bullet$, rather than single mass perturbers. In this case, S2 precession and inclination probe different moments of the mass distribution. On one hand, the coherent precession is set by the total enclosed mass and constrains $\langle m_\bullet\rangle\,n_0$, independently of the spectrum's shape. On the other hand,  the granular inclination random walk constrains the second moment $\langle m_\bullet^2\rangle\,n_0$
and is weighted toward the heavy tail. Our monochromatic limits therefore carry
over to any spectrum upon reading the precession bound at the mean mass
$\langle m_\bullet\rangle$ and the granular bound at the effective mass
$m_{\rm eff}=\langle m_\bullet^2\rangle/\langle m_\bullet\rangle$, with the
latter becoming the relevant limit only for a markedly top-heavy population. Importantly,
this has little impact on our final predictions, because the constraining limit over most of the parameter space is precession, which fixes the total mass and yields
$n_0\propto m_\bullet^{-1}$, approximately cancelling the explicit mass dependence of
the rates. As a result, we will see that our EMRI and EMRB predictions are largely
insensitive to the detailed shape of the sBH mass function at fixed total mass, which further justifies the monochromatic treatment adopted throughout.

\section{Scaling GRAVITY results to other galaxies} 
\label{sec:scaling}

\paragraph{Stellar-mass BH profile.} The constraints of Sec.~\ref{sec:gravity} calibrate the sBH population around a single
nucleus, Sgr~A$^*$ with $M_\bullet \simeq 4\times10^{6}\,M_\odot$. However, in the following we would like to compute also the extragalactic rates and associated SGWBs for both EMRBs and EMRI. To do so, one needs to
sum over galaxies with central black holes spanning $M_\bullet = 10^{4}$--$10^{9}\,M_\odot$, and this requires a prescription for how the normalisation $n_0$, defined above at the fixed reference radius $r_0 = 200\,$AU, carries over to nuclei of different mass.

A physically motivated scaling follows from assuming that each nucleus hosts a relaxed cusp anchored at its own sphere of influence. The influence radius, $r_h = GM_\bullet/\sigma^2$, encloses a stellar mass $\simeq 2M_\bullet$ by definition, and the $M_\bullet$--$\sigma$ relation, $M_\bullet\propto\sigma^{4}$~\cite{Tremaine:2002js} (here $\sigma$ is the velocity dispersion sigma of the host galaxy), gives $r_h\propto M_\bullet^{1/2}$.  Assuming for simplicity that a fixed fraction of the stars leave sBH remnants, the number of sBHs inside the sphere of influence scales with the stellar mass there, $N_\bullet(<r_h)\propto M_\bullet$. Integrating the density profile $n_\bullet(a) = n_0\,(a_0/a)^{\gamma}$ over the sphere
of influence and solving for the normalisation,
\begin{equation}
    N_\bullet(<R_h)
    =
    \frac{4\pi\, n_0\, a_0^{\gamma}}{3-\gamma}\,r_h^{3-\gamma}
    \Longrightarrow
    n_0 \propto\ \frac{M_\bullet}{r_h^{3-\gamma}}
    \propto
    M_\bullet^{(\gamma-1)/2},
    \label{eq:n0_scaling}
\end{equation}
for $\gamma < 3$, i.e.\ $n_0\propto M_\bullet^{3/8}$ for the BW slope. The weak, but positive exponent reflects the fact that a heavier black hole binds proportionally more remnants, but also spreads them over a much larger influence volume, $r_h^{3-\gamma}\propto M_\bullet^{5/8}$.
Note that Eq.~\eqref{eq:n0_scaling} keeps the enclosed cusp mass a fixed fraction of the central mass, $M_{\rm cusp}(<r_h)/M_\bullet \simeq 1/3$ when saturating the GRAVITY bound, whereas a mass-independent $n_0$ would imply a remnant cluster \emph{outweighing}
the central black hole for $M_\bullet\lesssim 2\times10^{5}\,M_\odot$, which is precisely the mass range that will dominate our extragalactic integrals.
Notice that the mass fraction implied by the upper limits by GRVITY is $M_{\rm sBH}/M_\bullet\lesssim0.14$ within 1 pc ($\approx0.34$ within $r_h$), corresponding to a few percent of the stellar mass ending up in sBH remnants. A standard initial mass function converts a fraction $\sim10^{-3}$ of stars into $\sim10M_\odot$ remnants~\cite{2001MNRAS.322..231K}, which then mass segregation concentrates toward Sgr A*~\cite{Hopman_2006}, boosting the local abundance by about an order of magnitude. The implied count at the S2 ceiling, $N_\bullet(<1{\rm pc})\lesssim6\times10^4$, is then within a factor of a few of the mass-segregated population-synthesis expectation, $N_\bullet(<1{\rm pc}) =(1$–$4)\times10^4$~\cite{MiraldaEscude2000ApJ, 2006ApJ...649...91F}. This shows that the sensitivity of GRAVITY is reaching typical expectations for the sBH density, with an astrometric measurement replacing the current upper limit on $n_0$ within reach.

Interestingly, this simple scaling argument seems validated by the numerical loss-cone calculations of Ref.~\cite{Hopman:2009}, where the author finds an EMRI rate scaling as $\Gamma_{\rm EMRI}\propto M_\bullet^{-1/4}$. As we will show below, assuming a BW profile our per-galaxy rate obeys
$\Gamma_{\rm EMRI}\propto n_0^{6/5}\,M_\bullet^{-7/10}$
[Eq.~\eqref{eq:emri_rate_BW_analytic} together with
$a_c\propto (M_\bullet/n_0)^{4/5}$], so that a power law $n_0\propto M_\bullet^{p}$
predicts $\Gamma_{\rm EMRI}\propto M_\bullet^{6p/5-7/10}$. Matching the simulated slope yields $p = 3/8$, in agreement with the value derived above, so that the simulations independently select the self-similar scaling of Eq.~\eqref{eq:n0_scaling}. In the following we therefore adopt
\begin{equation}
n_0(M_\bullet) \leq n_0^{\max}\,
    \left(\frac{M_\bullet}{4\times10^{6}\,M_\odot}\right)^{3/8},
    \label{eq:n0_prescription}
\end{equation}
with $n_0^{\max}(\gamma=7/4, m_\bullet)$ the GRAVITY limit of Sec.~\ref{sec:gravity}, as our fiducial calibrated extrapolation.
In the remainder of this work, whenever the normalisation $n_0$ appears without an explicit argument, the
mass-dependent normalisation $n_0(M_\bullet)$ of Eq.~\eqref{eq:n0_prescription} will always understood; we suppress the argument purely to lighten the notation.

We notice that the self-similar prescription of Eq.~\eqref{eq:n0_prescription} is possibly conservative at the low MBH mass end (which dominates the GW signal for LISA), as it presumes the same remnant abundance per unit stellar mass in every
nucleus. However, galactic nuclei could sit in the \emph{strong} mass-segregation regime, concentrating the remnants at the small radii relevant here and enhancing the corresponding densities by up to an order of magnitude~\cite{Alexander:2008tq,Preto:2009kd}. This effect is most relevant precisely for the light nuclei dominating the extragalactic EMRB signal, because their relaxation times are far shorter than a Hubble time, so
their cusps are most possibly fully relaxed and segregated. To bracket uncertainties, one could also use \emph{maximal} normalisation
consistent with both the GRAVITY anchor and the requirement that the cusp not outweigh the central black hole,
\begin{equation}
    n^\text{max}_0(M_\bullet)
    =
    n_0^{\max}\,
    \min\,\left[1,\left(\frac{M_\bullet}{M_{\rm crit}}\right)^{3/8}\right],\,
   \label{eq:n0_ceiling}
\end{equation}
with $ M_{\rm crit}\simeq 2\times10^{5}\,M_\odot$,
i.e.\ a mass-independent normalisation down to the mass $M_{\rm crit}$ at which the enclosed cusp mass reaches $M_\bullet$; below $M_{\rm crit}$ the normalisation is instead fixed by holding the cusp at that maximal mass,
$M_{\rm cusp}(<r_h)=M_\bullet$, which again implies $n_0\propto M_\bullet^{3/8}$ but
with the largest normalisation the mass budget allows. We stress that Eq.~\eqref{eq:n0_ceiling} is a purely phenomenological scaling, and serves to illustrate the uncertainties arising from the extrapolation of $n_0$ as measured by GRAVITY to extragalactic sources.

Finally, we note that the hierarchy between $r_I$ and $r_h$ discussed at the end of the previous section is preserved under the self-similar scaling of Eq.~\eqref{eq:n0_prescription}, so BW remains the appropriate fiducial profile across the extragalactic range. This is true also for the less conservative scaling~\eqref{eq:n0_ceiling}, which implies a larger $n_0$ and hence a larger sBH  to star fraction $f_\bullet$.

\paragraph{MBH mass function}

To compute the extragalactic rate of EMRB and EMRI events we need to make assumptions on the MBH distribution in our Universe. To this end, we use the local, occupation-corrected central black-hole mass functions inferred in Ref.~\cite{Burke:2024wcf}.
These mass functions are obtained by combining constraints on the black-hole occupation fraction in nearby galaxies with the local galaxy stellar-mass function and the $M_{\rm BH}-M_\star$ relation.  They therefore aim to describe the total central black-hole population, including inactive black holes, rather than only the active AGN population.  We digitize representative upper,
central, and lower $1\sigma$ curves from their Fig.~12 and use them as optimistic, fiducial, and pessimistic scenarios for $dn_{\rm BH}/d\ln M_{\rm BH}$. We show these mass functions in Fig.~\ref{fig:bhmf}.

\begin{figure}
    \centering
    \includegraphics[width=\linewidth]{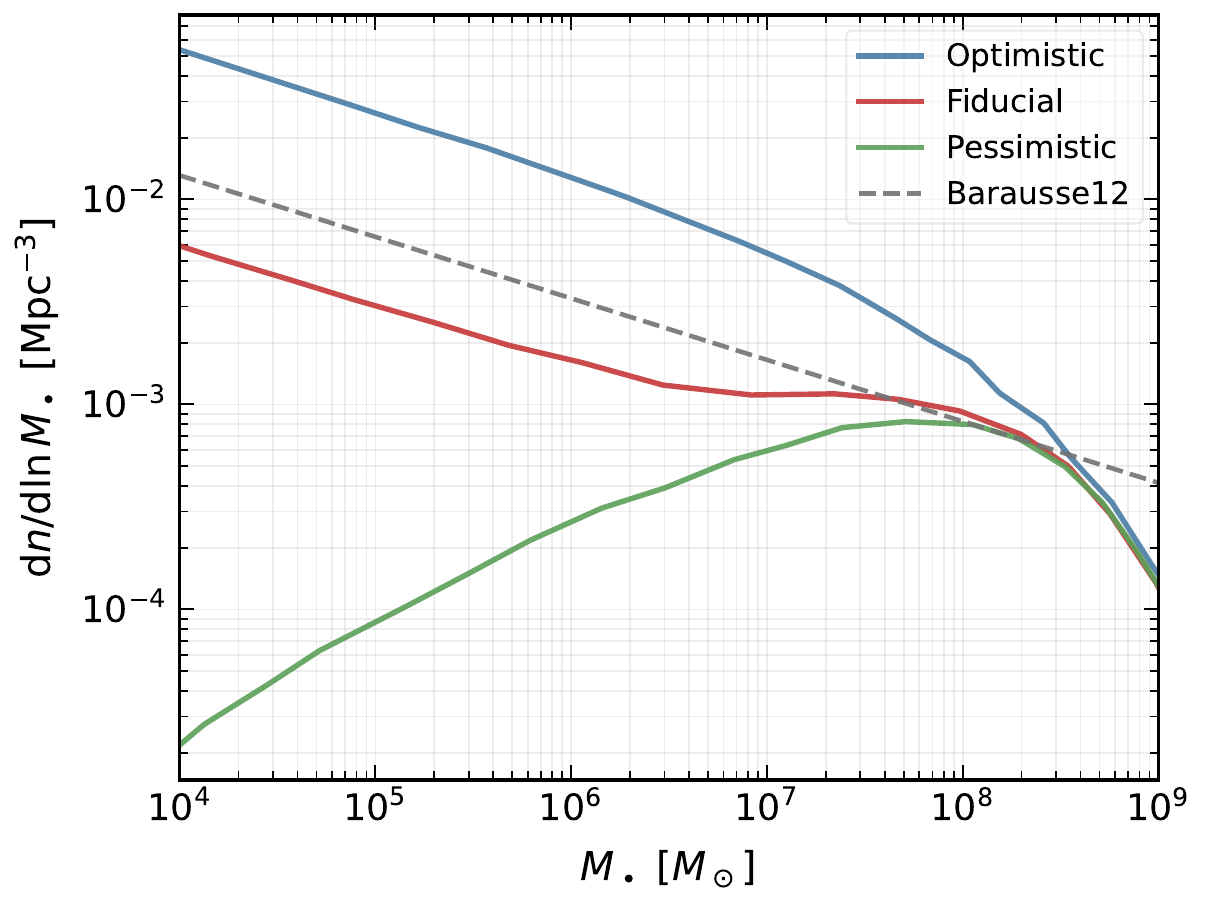}
    \caption{Local, occupation-corrected central black-hole mass function $dn/d\ln M_\bullet$ (in $\mathrm{Mpc}^{-3}$) as a function of black-hole mass $M_{\rm BH}$, for the three scenarios adopted in this work: optimistic (blue), fiducial (red), and pessimistic (green). The curves are digitized from the representative upper, central, and lower mass functions of Ref.~\cite{Burke:2024wcf} (their Fig.~12). The dashed gray line is the power-law fit $dn/d\ln M_\bullet = 0.005 (M_\bullet/4\times10^6 M_\odot)^{-0.3}\rm Mpc^{-3}$  of Ref.~\cite{Babak:2017tow} to the local mass function of the semi-analytic model of Ref.~\cite{Barausse:2012fy}.}
    \label{fig:bhmf}
\end{figure}

In the following, we will  present results for a fiducial, as well as an enhanced (optimistic) and suppressed (pessimistic) mass function in this range. We notice that the reliability of these mass functions is strongly mass dependent. They are most robust at the high-mass end, $M_{\rm BH}\gtrsim 10^{7}\,M_\odot$, where the $M_{\rm BH}$--$M_\star$ relation is
directly calibrated and the residual uncertainty is set mainly by its intrinsic scatter. The low-mass end is instead governed by the \emph{slope} of the $M_{\rm BH}$--$M_\star$ relation and by the shape of the occupation fraction, which
Ref.~\cite{Burke:2024wcf} constrains observationally only down to host stellar masses $M_\star\sim 10^{7}$--$10^{8}\,M_\odot$. Accordingly, those authors explicitly caution that their mass function is very uncertain below $M_{\rm BH}\sim 10^{5}\,M_\odot$, where it relies on
extrapolating the $M_{\rm BH}$--$M_\star$ relation beyond its calibration range, and that it becomes a very uncertain extrapolation below $M_{\rm BH}\sim 10^{4}\,M_\odot$. It is precisely in this intermediate-mass regime that the three curves in Fig.~\ref{fig:bhmf} diverge by up
to two--three orders of magnitude; the spread between the optimistic, fiducial, and pessimistic scenarios at $M_{\rm BH}\lesssim 10^{5}\,M_\odot$ should thus be
interpreted as the (large) systematic uncertainty on the occupation fraction rather than as a statistical error, and it is a dominant source of uncertainty in our
low-mass EMRI/EMRB host population. To our knowledge, this is the first time the occupation-corrected local black-hole mass
function of Ref.~\cite{Burke:2024wcf} has been propagated to the extragalactic EMRI and EMRB
gravitational-wave backgrounds. Previous background estimates~\cite{Berry:2013ara,Bonetti:2020jku,Pozzoli:2023kxy} relied on the semi-analytic or power-law
mass functions of Refs.~\cite{Barausse:2012fy,Babak:2017tow}, which do not incorporate the
observationally constrained occupation fraction that shapes the low-mass end most relevant for these sources.

For comparison, in Fig.~\ref{fig:bhmf} we also show the single power-law fit $dn/d\ln M_\bullet = 0.005 (M_\bullet/4\times10^6 M_\odot)^{-0.3}\rm Mpc^{-3}$  of Ref.~\cite{Babak:2017tow} to the local mass function of the semi-analytic model of Ref.~\cite{Barausse:2012fy}, adopted also in Ref.~\cite{Rom:2024nso}. Notice that this power-law fit to $\sim 2\times$ our fiducial at $\lesssim 10^5 M_\odot$, while our optimistic is $\sim 4 \times$ above that. So it sits between our fiducial and optimistic choices.

\section{Rates and gravitational-wave background from compact-objects around a MBH.}
\label{sec:general_formalism}

This section summarizes the key ingredients to quantify GW signatures from EMRBs and EMRIs, in particular event rates and stochastic backgrounds, based on a series of earlier works~\cite{Hopman:2005vr, Berry:2013ara, Babak:2017tow, Bonetti:2020jku, Rom:2024nso}. Our focus is on presenting a unified framework for consistent predictions across different GW signatures which are based on the same underlying MBH and sBH source populations.

\paragraph{Black hole phase space.} Stellar-mass black holes orbiting an MBH give rise to several related gravitational-wave phenomena,  which originate from different parts of the same underlying phase-space distribution of $F_\bullet$ of stellar-mass remnants. In the following we review the relevant phase space for EMRBs, EMRIs and direct plunges following he pedagogical description given in Ref.~\cite{Rom:2024nso,Berry:2013ara}.
%
%
%
For a Keplerian bound orbit the periapse $r_p$ is related to the semimajor axis $a$ by $r_p=a(1-e)$, where $e$ is the eccentricity.  We also define the Schwarzschild radius
\begin{equation}
    r_s\equiv \frac{2G M_\bullet}{c^2},
\end{equation}
and denote the capture boundary by 
\begin{equation}
\label{eq:rcapture}
    r_{\rm p, cap}\simeq 4r_s,
\end{equation}
for nearly parabolic orbits around a non-spinning MBH.

\begin{figure}
    \centering
    \includegraphics[width=1.02\linewidth]{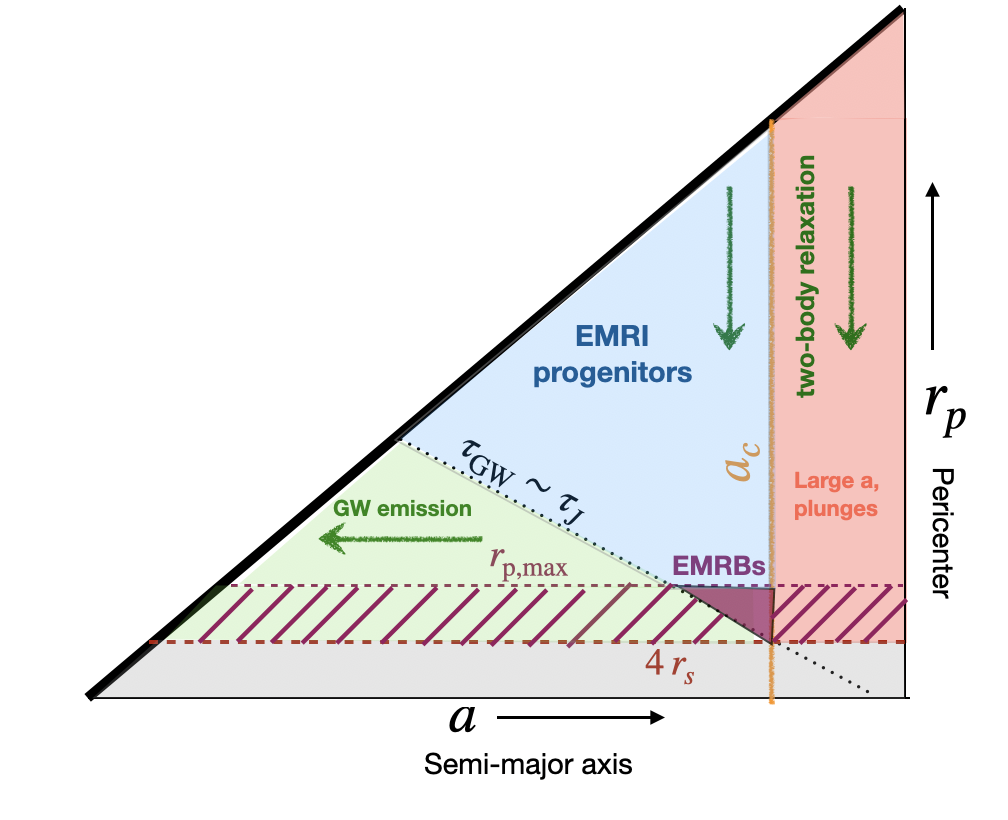}
    \caption{Schematic phase-space structure in the $(a,r_p)$ plane for compact objects orbiting the central MBH, inspired by Fig.~1 of Ref.~\cite{Rom:2024nso}.  The upper diagonal black line indicates the physical boundary $r_p=a$.
    The dotted black curve, $\tau_{\rm GW}\simeq \tau_{J}$, separates the scattering-dominated evolution of EMRI progenitors (blue region) from GW-dominated inspiral (green region). The gray region at small $r_p < 4 r_s$ indicated mostly bound orbits.  At large $a$, two-body relaxation drives objects directly toward the capture boundary, producing plunges rather than EMRIs (red region). The green arrows give an indication of the typical trajectories of binaries in this phase space, driven by two-body scattering and GW emission.
    The horizontal dashed line marks the maximum periapsis $r_{p,\max}$ for detectable bursts, identifying the purple region as the phase-space region contributing to observable EMRBs. The hatched region corresponds to phase space which would naively satisfy the burst periapsis condition, but exhibits a significantly suppressed number density compared to the quasi-stationary distribution maintained by two-body sacttering. This is because these orbits
    are either depleted by GW emission faster than relaxation can refill them, or are better interpreted as direct plunges.}
    \label{fig:phasespace}
\end{figure}

The competition between gravitational-wave emission and two-body relaxation is
described by the two timescales
\begin{equation}
    \tau_{\rm GW}(a,r_p),
    \qquad
    \tau_J(a,r_p),
\end{equation}
 where $\tau_J$ is the angular-momentum relaxation time, which is the typical time scale at which compact objects are scattered in to or out of a given orbit. For $\Mb \gg \mn$, the orbit-averaged gravitational-wave inspiral time obtained from the Peters energy-loss formula is~\cite{Peters:1963ux, Berry:2013ara}
\begin{equation} \tau_{\rm GW}(a,r_p) = \frac{5}{64} \frac{c^5 r_p^4}{G^3 \Mb^2\mn} \frac{  \left[ \frac{a}{r_p}\right]^{1/2} \left[ \left(2-\frac{r_p}{a}\right) \right]^{7/2}} {1+\frac{73}{24}\left(1-\frac{r_p}{a}\right)^2 +\frac{37}{96}\left(1-\frac{r_p}{a}\right)^4},
\label{eq:tauGW-arp} 
\end{equation}
while the angular-momentum relaxation time is estimated as~\cite{1987degc.book.....S,binney1987}  
\begin{equation} \tau_{J}(a) = \frac{0.34\,\sigma(a)^3(1-e^2)} {G^2m_\bullet^2 n(a)\ln\Lambda},
\label{eq:taur-def} 
\end{equation} 
where $\ln\Lambda \sim \ln\Big(M_\bullet/m_\bullet\Big)$ is the the relaxation Coulomb logarithm, and $\sigma$ is the dispersion velocity, for which we use the Keplerian estimate \begin{equation} \sigma(a) = \left[\frac{GM_\bullet}{(1+\gamma)a}\right]^{1/2}.
\label{eq:sigma-kepler} 
\end{equation}

The curve $\tau_{\rm GW}(a,r_p)\simeq\tau_J(a,r_p)$ separates scattering-dominated and GW-dominated evolution.  We then define
$a_c$ by
\begin{equation}
\label{eq:GW2b}
    \tau_{\rm GW}\!\left(a_c,r_{\rm p, cap}\right)
    =
    k\,\tau_J\left(a_c,r_{\rm p, cap}\right),
\end{equation}
indicated by the dotted oblique black line in Fig.~\ref{fig:phasespace}. Here $k\simeq 1.36$ is an order-unity factor obtained from solving a Fokker-Planck equation in energy-angular moment space~\cite{Kaur:2024ofj}. It separates the phase space of EMRI progenitors, evolving mainly under 2-body scattering, from the EMRIs, evolving mainly under GW emission. Objects with $a<a_c$ can enter the GW-dominated region before crossing the
direct-capture boundary, and therefore can become EMRIs.  Objects with
$a>a_c$ generally reach $r_p\simeq r_{\rm p, cap}$ before GW dissipation can
decouple them from relaxation, and therefore plunge directly. This sharp division at $a_c$ is, however, an idealisation. More detailed
treatments of the inspiral--plunge boundary find that successful captures can also originate from semi-major axes beyond
$a_c$, the so-called ``cliffhanger'' EMRIs~\cite{Qunbar:2023vys, Mancieri:2024sfy}, which narrowly avoid direct plunge. By truncating the EMRI-forming region at $a_c$ we neglect this channel. 

The different parts of the $r_p - a$ phase space are illustrated in Fig.~\ref{fig:phasespace}.

\paragraph{Event rates.} For any source class $X$, we write the rate in a single galaxy as
\begin{equation}
    \Gamma_X(M_\bullet)
    =
    \int_{\mathcal D_X}
    d\boldsymbol{\theta}_X\,
    \mathcal K_X(M_\bullet,\boldsymbol{\theta}_X)\,
    \mathcal W_X(M_\bullet,\boldsymbol{\theta}_X).
    \label{eq:general_rate}
\end{equation}
Here $\boldsymbol{\theta}_X$ denotes the orbital variables appropriate to the
source class, $\mathcal D_X$ is the selected region of phase space,
$\mathcal K_X$ is the corresponding rate kernel, and $\mathcal W_X$ is an
selection or depletion factor (which will be specified later). For the cases of interest, $\boldsymbol{\theta}_X$ will consist of two orbital variables at most, which will be chosen among eccentricity, pericenter and semi-major axis.

The same parent compact-object distribution enters all source classes, but the kernels and integration domains are different, and the distinction is crucial.  An EMRB is an individual close periapse passage.  It is therefore counted with a passage-rate kernel.  An EMRI, by contrast, is a successful GW capture into a long-lived inspiral.  It is
therefore counted with a flux into the GW-dominated capture region.  Thus EMRBs
and EMRIs can be written in the same schematic form, Eq.~\eqref{eq:general_rate},
but they do not correspond to the same microscopic event.

Equation~\eqref{eq:general_rate} defines the rate of source class $X$ in a
single galaxy hosting an MBH of mass $M_\bullet$.  To compute either the
cosmological event rate or the associated stochastic gravitational-wave
background, this rate must be convolved with the cosmological population of
MBHs.  We denote the comoving MBH mass function by
\begin{equation}
    \Phi(M_\bullet,z)
    \equiv
    \frac{dn_{\rm MBH}}{d\ln M_\bullet}(M_\bullet,z),
    \label{eq:MBH_mass_function}
\end{equation}
where $\Phi$ has units of comoving number density.  This quantity should
not be confused with the compact-object number density $n_\bullet(r)$ around
an individual MBH.  The latter determines the rate kernel
$\mathcal K_X$, while $\Phi$ determines how many such galactic nuclei
exist per comoving volume. The cosmological expansion is described by
\begin{equation}
    H(z)=H_0 E(z),
    \qquad
    E(z)=
    \left[
    \Omega_m(1+z)^3+\Omega_\Lambda
    \right]^{1/2},
    \label{eq:Ez}
\end{equation}
the comoving distance is
\begin{equation}
    D_c(z)
    =
    c\int_0^z \frac{dz'}{H(z')},
\end{equation}
and the comoving volume element per unit redshift and solid angle
\begin{equation}
    \frac{dV_c}{dz\,d\Omega}
    =
    \frac{cD_c^2(z)}{H(z)},
\end{equation}
or, after integrating over the full sky,
\begin{equation}
    \frac{dV_c}{dz}
    =
    4\pi
    \frac{cD_c^2(z)}{H(z)} .
    \label{eq:comoving_volume}
\end{equation}

The differential observed event rate from source class $X$ is then
\begin{eqnarray}
    \frac{d^3\dot N_X}
    {dz\,d\ln M_\bullet\,d\boldsymbol{\theta}_X}
    =&
    \frac{1}{1+z}
    \frac{dV_c}{dz}
    \Phi(M_\bullet,z)
    \mathcal K_X(M_\bullet,\boldsymbol{\theta}_X,z)\nonumber\\
    &\times\mathcal W_X(M_\bullet,\boldsymbol{\theta}_X,z),
    \label{eq:observed_event_rate_differential}
\end{eqnarray}
where the factor $(1+z)^{-1}$ converts source-frame rates into observer-frame rates.
Integrating over the phase-space domain gives
\begin{equation}
    \frac{d^2\dot N_X}
    {dz\,d\ln M_\bullet}
    =
    \frac{1}{1+z}
    \frac{dV_c}{dz}
    \Phi(M_\bullet,z)
    \Gamma_X(M_\bullet,z),
    \label{eq:observed_event_rate_mass_redshift}
\end{equation}
with $\Gamma_X$ defined in Eq.~\eqref{eq:general_rate}.  Finally, the total
observed event rate is
\begin{equation}
    \dot N_X
    =
    \int dz
    \int d\ln M_\bullet\,
    \frac{1}{1+z}
    \frac{dV_c}{dz}
    \Phi(M_\bullet,z)
    \Gamma_X(M_\bullet,z).
    \label{eq:observed_total_event_rate}
\end{equation}
This expression is useful, for example, when estimating the number of
detectable extragalactic EMRBs or the all-sky rate of cosmological
bursts, whereas for the galactic rate Eq.~\eqref{eq:general_rate} suffices.

\paragraph{Stochastic gravitational wave background.} To compute the associated stochastic gravitational wave background, we need to attach the gravitational-wave energy emitted by each event.  For a
source class $X$, let
\begin{equation}
    \frac{dE_X}{df_s}
    =
    \frac{dE_X}{df_s}
    \left(
    f_s;
    M_\bullet,m,\boldsymbol{\theta}_X
    \right)
    \label{eq:single_source_spectrum_general}
\end{equation}
be the source-frame GW energy spectrum of one event, where
\begin{equation}
    f_s=(1+z)f_{\rm obs}
\end{equation}
is the source-frame frequency corresponding to the observed frequency
$f_{\rm obs}$.  For EMRBs, $dE_X/df_s$ is the spectrum of a single close
periapse burst, while for EMRIs, it is the spectrum emitted during the subsequent
inspiral associated with one capture event. 

The stochastic background is defined by
\begin{equation}
    \Omega_{\rm GW}^{X}(f_{\rm obs})
    \equiv
    \frac{1}{\rho_c}
    \frac{d\rho_{\rm GW}^{X}}{d\ln f_{\rm obs}},
    \label{eq:omega_definition}
\end{equation}
with $\rho_c$ the critical energy density,
\begin{equation}
    \rho_c
    =
    \frac{3H_0^2c^2}{8\pi G}.
    \label{eq:critical_energy_density}
\end{equation}
With this convention, the contribution from source class $X$ to the GW-background can be written as~\cite{Phinney:2001di}
\begin{align}
    & \Omega_{\rm GW}^{X}(f_{\rm obs})
    =
    \frac{f_{\rm obs}}{\rho_c H_0}
    \int dz\,
    \frac{1}{(1+z)E(z)}
    \int d\ln M_\bullet\,
    \Phi(M_\bullet,z) \nonumber\\
    & \;\; \int_{\mathcal D_X}
    d\boldsymbol{\theta}_X\,
    \mathcal K_X(M_\bullet,\boldsymbol{\theta}_X,z)
    \mathcal W_X(M_\bullet,\boldsymbol{\theta}_X,z)
    \left.
    \frac{dE_X}{df_s}
    \right|_{f_s} \!\!. 
    \label{eq:omega_general}
\end{align}
The astrometric input enters through the compact-object distribution around
each MBH, determining the strength of the EMRB and EMRI source kernels inside each galactic nucleus. The MBH mass function $\Phi$ determines instead the number of galactic nuclei contributing to the cosmological integral.

Equations~\eqref{eq:observed_total_event_rate} and
\eqref{eq:omega_general} are the general expressions that below we will apply concretely to EMRBs and EMRIs below.
For EMRBs, 
the kernel counts close periapse passages, and the spectrum is the energy
emitted in one burst.  For EMRIs, the kernel counts successful GW captures,
and the spectrum is that of the long-lived inspiral following capture.  The
two source classes can therefore be treated within the same cosmological
framework, but with different phase-space domains, rate kernels, and single-source GW spectra.

\subsection{EMRBs} For EMRBs, the natural variables are $\boldsymbol{\theta}_{\rm B}=(e,r_p)$. An EMRB is produced when a compact object passes close to the MBH, with
\begin{equation}
    r_{\rm p, cap}<r_p<r_{p,\max};
\end{equation}
the lower limit removes direct plunges, while the upper limit
is set by the observation threshold. In particular, for a detectable Galactic EMRB rate, $r_{p,\max}$ is set by the sensitivity of the observing instrument, while in the case of the stochastic background, it is set by the observed frequency range. This definition of EMRBs includes what has been referred to as fly-bys~\cite{Toonen:2009qi} or EMRI peeps~\cite{Oliver:2023xan} in the literature. The latter refer to repeated GW bursts from a sBH on a highly eccentric orbit with a large orbital period. For the purpose of our discussion, there is no need to distinguish these cases, as we it will suffice to focus on the near parabolic orbit at periapse passage.

For an isotropic compact-object distribution function $F_\bullet$, the
differential EMRB passage rate can be written as~\cite{Berry:2013ara}
\begin{equation}
    \frac{d^2\Gamma_{\rm B}}{de\,dr_p}
    =
    (2\pi)^2
    (G M_\bullet)^2
    \frac{e}{r_p}
    F_\bullet(\varepsilon),
    \label{eq:emrb_rate_kernel}
\end{equation}
where
\begin{equation}
F_\bullet(\varepsilon) = \frac{\Gamma[\gamma+1]\varepsilon^{\gamma-3/2}}{2\sqrt{2}\,\pi^{3/2}\,\Gamma[\gamma-1/2]}
\left(\frac{a_0}{GM_\bullet}\right)^{\gamma} n(a_0)\,
\end{equation}
with $\varepsilon=\frac{G M_\bullet}{2a}$ being the binding energy per unit mass. For a full derivation of this (known) result the interested reader can see App.~\ref{app:eddington}. The total EMRB rate in one
galaxy is therefore
\begin{equation}
    \Gamma_{\rm B}(M_\bullet)
    =
    \int_{r_{\rm p, cap}}^{r_{p,\max}}
    dr_p
    \int_{e_{\min}}^{1}
    de\,
    \mathcal W_{\rm B}(e,r_p)
    \frac{d^2\Gamma_{\rm B}}{de\,dr_p}.
    \label{eq:total_emrb_rate}
\end{equation}

The eccentricity lower limit specifies which close passages are treated as
bursts, and we adopt 
\begin{equation}
    e_{\min}=0.9,
\end{equation}
as our fiducial choice.  A useful measure of the accuracy of this
threshold is the spacing of the discrete Peters--Mathews harmonics relative to the
characteristic burst frequency.  Since
$f_{\rm orb} = (1-e)^{3/2} f_c$,  the dimensionless harmonic spacing is
$\frac{\Delta f}{f_c}=(1-e)^{3/2}$. For  $e_{\min}=0.9$, one has $(1-e_{\min})^{3/2}\simeq 3.2\times 10^{-2}$, so that the spectrum is already well approximated by a continuum, i.e.\ better described by a burst than a few harmonics. Smaller values of $e_\text{min}$ would increase our rate predictions, at the cost of a loss of accuracy when employing parabolic approximations. Note also that $r_{p,\text{max}}$ implies a lower limit on $e$ (see purple region in Fig.~\ref{fig:phasespace}), since the GW signal is suppressed at low eccentricities.

Once binaries enter the region dominated by GW emission, their number density is rapidly depleted compared to a steady-state profile maintained by two body scattering. We model this by adopting a smooth depletion description encoded in $\mathcal W_{\rm B}$\footnote{Suppose the Bahcall--Wolf cusp (or any other cusp) would like to maintain some phase-space occupation $f_{\mathrm{BW}}$, while two-body scattering tries to refill a given region of phase space on a timescale $\tau_J$, and gravitational-wave inspiral drains it on a timescale $\tau_{\mathrm{GW}}$. A simple toy evolution equation is then
\begin{equation}
\frac{df}{dt}
=
\frac{f_{\mathrm{BW}}-f}{\tau_J}
-
\frac{f}{\tau_{\mathrm{GW}}}.
\end{equation}
The first term describes relaxation toward the unperturbed occupation $f_{\mathrm{BW}}$ on the replenishment timescale $\tau_J$, while the second term accounts for depletion by GW inspiral. In steady state one sets $\frac{df}{dt}=0$, and derive the steady-state occupation 
\begin{equation}
f
=
f_{\mathrm{BW}}
\frac{\tau_{\mathrm{GW}}}{\tau_{\mathrm{GW}}+\tau_J},
\end{equation}
which suggests the smooth suppression factor $\mathcal W_{\rm B}=\frac{\tau_{\rm GW}}
    {\tau_{\rm GW}+\tau_J}$ we adopt.},
\begin{equation}
    \mathcal W_{\rm B}
    =
    \frac{\tau_{\rm GW}}
    {\tau_{\rm GW}+\tau_J};
    \label{eq:depletion_weight}
\end{equation}
which approaches unity in the scattering-dominated regime,
$\tau_{\rm GW}\gg\tau_J$, and suppresses the GW-dominated region,
$\tau_{\rm GW}\ll\tau_J$, where compact objects are depleted by capture into EMRIs. This is the reason why the EMRBs region in Fig.~2 is hatched to the right of the dotted line denoting where $\tau_{\rm GW} \simeq \tau_{J}$.

We note that the differential rate $d^2 \Gamma_B/de dr_p$ depends linearly on $n_0$, with a further dependence on $n_0$ entering through $\tau_J$ in the depletion factor. This results in resulting in a stronger than linear, but at most quadratic dependence on $n_0$ in the parameter space of interest. By direct integration we find $\Gamma_B \propto n_0^{\alpha}$ with a local exponent $\alpha \simeq 1.6$ at the S2-anchored density (the exponent varies a bit with density, because there is less depletion as the cusp densifies).

\subsection{EMRIs}

EMRIs are produced by the same compact-object population as EMRBs, but they
correspond to a different outcome of the loss-cone dynamics. An EMRB is an
individual close-periapse passage, whereas an EMRI is a successful capture into a long-lived GW-driven inspiral. The two phenomena therefore probe related regions of phase space, but they are counted differently, because EMRB rates count close passages, while EMRI rates count successful captures.

Formally, one may write the EMRI rate as an integral over the capture domain,
\begin{equation}
    \Gamma_{\rm I}(M_\bullet)
    =
    \int_{a_{\rm GW}}^{a_c} da
    \int_{4r_s}^{a}
    dr_p\,
    \mathcal W_{\rm I}(a,r_p)
    \frac{d^2\Gamma_{\rm I}}{da\,dr_p},
    \label{eq:total_emri_rate}
\end{equation}
where $a_{GW}$ is edge of the loss cone, $\tau_J \simeq \tau_{2B}$, at $e = 0$.
In practice, however, we use the standard empty-loss-cone approximation~\cite{Hopman:2005vr, 1977ApJ...211..244L, BarOr2016, Cohn1978ApJ}. At fixed $a$, the compact objects form a reservoir that diffuses in angular momentum, or equivalently in eccentricity. Solving this diffusion problem gives
the integrated flux into the low-angular-momentum region, with the eccentricity integral absorbed into the loss-cone flux. The reduced rate kernel is then
\begin{equation}
    \frac{d\Gamma_{\rm I}}{da}
    \simeq
    \frac{4\pi a^2 n(a)}{C_{\rm LC}\tau_J(a, e=0)},
    \label{eq:emri_reduced_kernel}
\end{equation}
where $C_{\rm LC}$ is the logarithmic loss-cone factor, which in our fiducial calculation we take to be $C_{\rm LC}=15$. This factor is different than the Coulomb logarithm $\ln\Lambda$ entering $\tau_J$.We note that one uses $\tau_J(a,e)$ to determine whether an already eccentric orbit decouples from relaxation, but then the rate itself is the integrated empty-loss-cone flux from the reservoir at fixed $a$, and is therefore controlled by $C_{\rm LC}\tau_J(a,e=0)$.

The intrinsic EMRI rate per galaxy is thus~\cite{Rom:2024nso}
\begin{equation}
    \Gamma_{\rm EMRI}(M_\bullet)
    = \int_{a_{\rm GW}}^{a_c}da\,
    \frac{4\pi a^2 n(a)}{C_{\rm LC}\tau_{\rm rlx}(a)} \,.
    \label{eq:emri_rate_per_galaxy}
\end{equation}
For our fiducial BW profile, the integrand is independent of $a$ and the per-galaxy rate simplifies to
\begin{equation}
\Gamma_{\rm EMRI}^{\rm BW}(M_\bullet)=\frac{4\pi\left(\frac{11}{4}\right)^{3/2}}{0.34\,C_{\rm LC}}
    \frac{G^{1/2}m_\bullet^2 n_0^2 r_0^{7/2}\ln\Lambda}
    {M_\bullet^{3/2}}\left(a_c-a_{\rm GW}\right).
    \label{eq:emri_rate_BW_analytic}
\end{equation}

We caution that although Eq.~\eqref{eq:emri_rate_BW_analytic} carries an explicit factor $n_0^2$ (one
power from the source density and one from the inverse relaxation time), the integration boundary $a_c$ is itself set by the cusp normalisation, and this softens the final scaling. Since $\tau_{\rm GW}\propto a^{1/2}$ while
$\tau_J\propto a^{\gamma-5/2}/(m_\bullet^2 n_0\ln\Lambda)$, one finds $a_c \propto \left(M_\bullet / (m_\bullet n_0 \ln\Lambda)\right)^{1/(3-\gamma)}$, i.e.\ $a_c\propto n_0^{-4/5}$ for the BW slope $\gamma=7/4$. A denser cusp relaxes faster, so scattering overwhelms gravitational-wave inspiral out to smaller radii and the EMRI-producing region shrinks. Since $a_{\rm GW}\ll a_c$, the rate is linear in $a_c$ and therefore scales as $\Gamma_{\rm EMRI}\propto n_0^{2-1/(3-\gamma)} = n_0^{6/5}$ rather than $n_0^2$. We have
verified this numerically over the full range of interest.


\section{Results}
\label{sec:results}

\subsection{EMRBs}


\paragraph{Event rate per galaxy.} To estimate the rate of detectable EMRBs in our galaxy, we start from Eq.~\eqref{eq:total_emrb_rate}, with the upper integration boundary $r_{p, \text{max}}$ set by the sensitivity of the observing instrument. For LISA, the signal-to-noise ratio (SNR) for these events can be estimated as~\cite{Berry:2012im}
\begin{align}
 \log \text{SNR} = - 2.7 \log \left( \frac{r_p}{R_s/2} \right) + \log \left( \frac{m}{M_\odot} \right) + 4.9 \,,
\end{align}
and we define $r_{p,\text{max}}$ to correspond to an SNR of 5.  Our fiducial choice for the distribution function of the compact objects entering Eq.~\eqref{eq:emrb_rate_kernel} is a Bahcall-Wolf profile, $\gamma = 7/4$, with a depletion factor due to GW emission given by Eq.~\eqref{eq:depletion_weight}.

The resulting predictions for the upper bound on the galactic EMRB rate are shown in
Figure~\ref{fig:EMRBrates}, with the sBH number density constrained by GRAVITY's observations as described in Sec.~\ref{sec:gravity}. Our results taking into account the depletion of the inner region of the sBH profile agree well with Ref.~\cite{Hopman:2006fc} which follows a similar procedure. They are significantly lower than previous estimates neglecting the depletion of the inner region of the sBH profile due to GW emission~\cite{Berry:2013ara}, as depicted by the dashed curve.
While at $2\sigma$, rates of ${\cal O}(1)$ EMRBs / year in our galaxy are still allowed, the preferred region predicts on average at most one EMRB per five years, making a detection within the mission duration of LISA less likely.
These predictions show only a mild dependence on the sBH mass, justifying the use of a monochromatic mass distribution for simplicity. Our derived limits for the galactic EMRB rate are in the ballpark of the estimates in Ref.~\cite{Hopman2007} and \cite{Berry:2013ara}, while they are a bit above than the rate estimates found in Ref.~\cite{Rubbo2006ApJ}. We attribute this to mass segregation which was not accounted for in Ref.~\cite{Rubbo2006ApJ}, see also discussion in Refs.~\cite{Hopman2007} and \cite{Berry:2013ara}.}

Uncertainties on the sBH mass profile in the inner-most region of our galaxy remain, and have a significant impact on these results. Upcoming astrometric data which could yield a measurement, instead of an upper bound for $n_0$, will significantly reduce these. We stress that for the galactic EMRBs, the observations by the GRAVITY collaboration are directly constraining the relevant source population, and hence immediately inform the expected rate in LISA.

\begin{figure}
    \centering
    \includegraphics[width=1.02\linewidth]{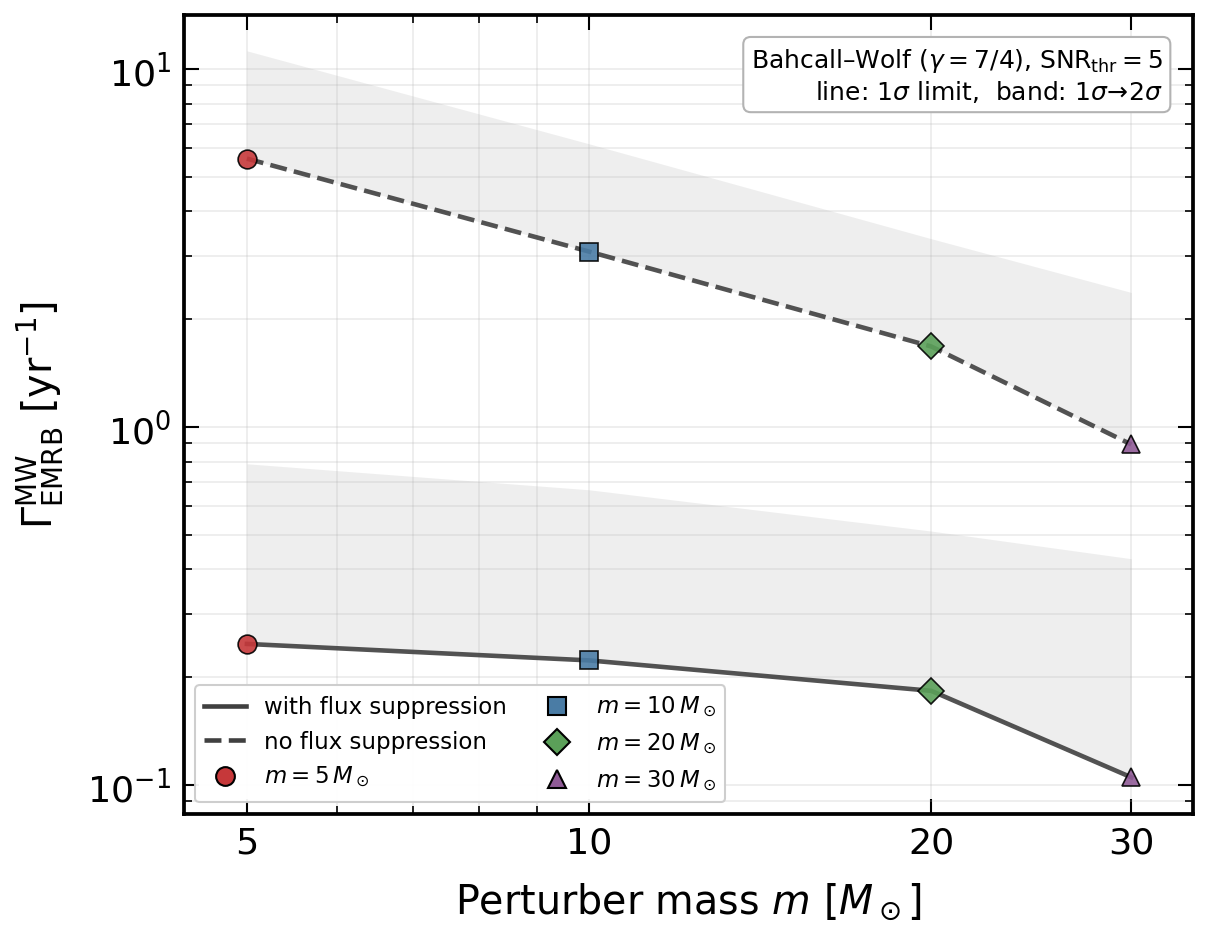}
    \caption{Expected rate of detectable extreme-mass-ratio bursts from Sgr~A$^*$, $\Gamma_{\rm EMRB}^{\rm MW}$, as a function of the stellar-mass black-hole mass $m$, for a Bahcall--Wolf cusp ($\gamma=7/4$) and a detection threshold ${\rm SNR}=5$. For
each $m$ the cusp normalisation $n_0$ is fixed to the strongest S2/GRAVITY upper limit (the binding $1\sigma$ bound; markers and colours as in Fig.~\ref{fig:n0_s2}). Solid and dashed curves give the rate with and without loss-cone (flux) suppression, and the shaded bands span the $1\sigma$ to $2\sigma$ GRAVITY confidence level.}
    \label{fig:EMRBrates}
\end{figure}

Going beyond the Milky Way, these numbers suggest that we can expect a significant amount of extragalactic EMRBs, which will constitute a SGWB. Contrary to the galactic EMRB rate, all following results will also depend on the MBH mass function and the extrapolation of the GRAVITY limit to other galaxies, introducing an additional systematic uncertainties. We estimate these by presenting results for the three different mass functions and $1-2\sigma$ error bands on $n_0^{\rm max}$.

\paragraph{SGWB.} To compute the spectrum of the gravitational wave background following Eq.~\eqref{eq:omega_general}, we additionally need to specify the energy emission per frequency in the source frame at peripase passage, $dE/df_s$,  and the MBH mass function $\Phi$. For the former, we use the Peters Matthews energy spectrum~\cite{Peters:1963ux} (which has been shown to agree well with more refined computations in~\cite{Berry:2012im} in the limit of high eccentricity relevant for EMRBs)~\cite{Berry:2010gt},
\begin{align}
 \frac{dE}{df} = \frac{4 \pi^2}{5} \frac{G^3 M_\bullet^2 \mu^2}{r_p^2} \ell(f/f_c) \,,
\end{align}
with $f_c = \sqrt{G M_\bullet/r_p^3}/(2 \pi)$ the orbital frequency and
\begin{align}
  \ell(\tilde f) = ( 8 \tilde f^2 B(\tilde f) - 2 \tilde f A(\tilde f))^2 + (128 \tilde f^4 + \frac{4 \tilde f^2}{3}) A^2(\tilde f) \,.
\end{align}
In App.~\ref{app:parabolic_spectrum} we give a quick derivation of energy spectrum and report the functions $A(\tilde{f})$ and  $B(\tilde{f})$ in Eq.~\ref{eq:A_parabolic}-\ref{eq:B_parabolic}.
The spectrum drops sharply at $f \gg f_c$, which strongly suppresses the integrand at large $r_p$. In practice, we limit the integration region to $f/f_c < 3.8$.

\begin{figure}
    \centering
    \includegraphics[width=\linewidth]{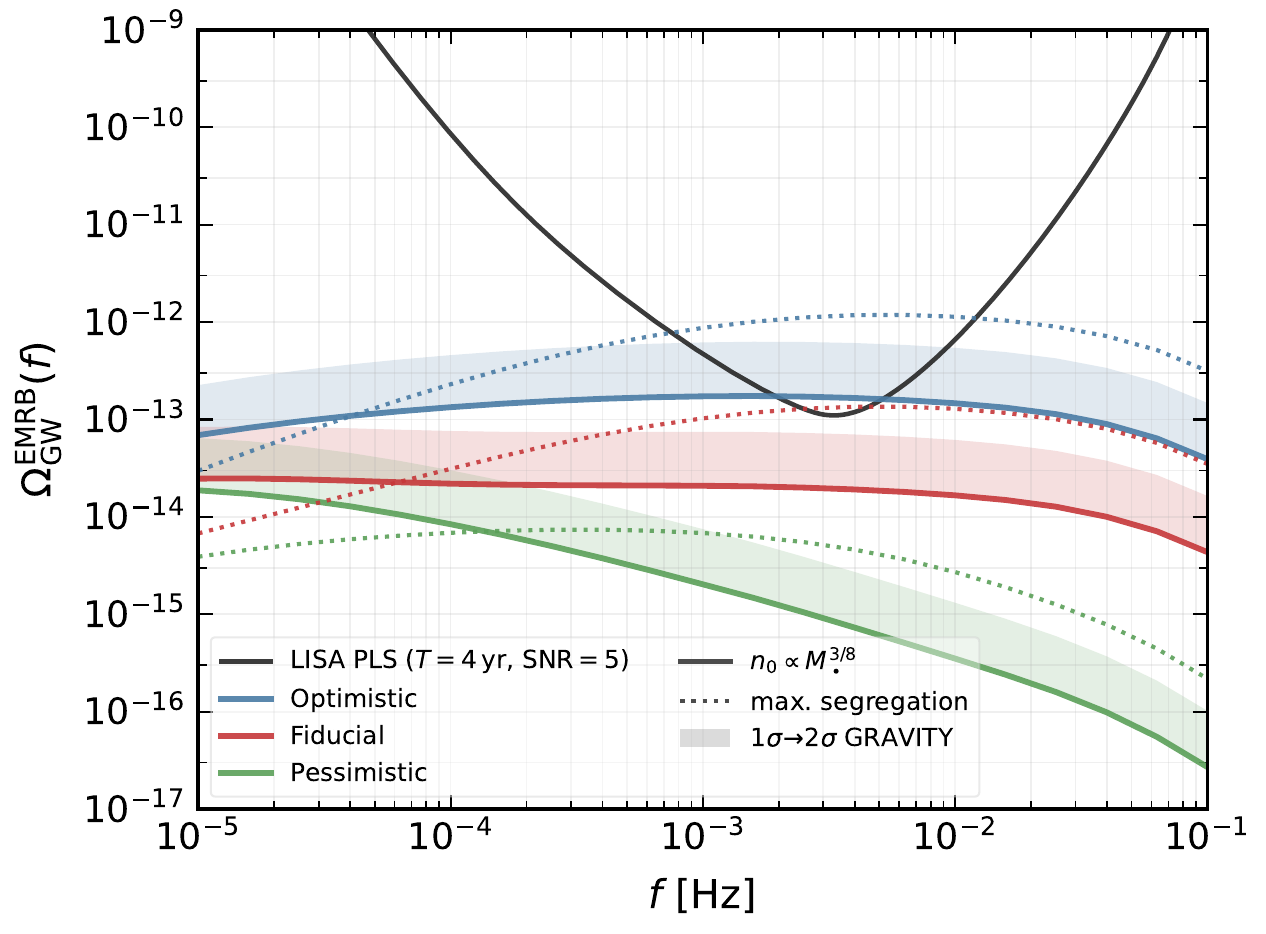}
    \caption{Stochastic gravitational wave background from the cosmological EMRB population, $\Omega_{\rm GW}^{\rm EMRB}(f)$, for the optimistic (blue), fiducial
(red), and pessimistic (green) central black-hole mass functions of Fig.~\ref{fig:bhmf}, computed with the S2/GRAVITY-constrained cusp normalisation for $m=10\,M_\odot$ perturbers. Solid lines adopt the self-similar prescription for the sBH number density, Eq.~\eqref{eq:n0_prescription}, with the shaded bands spanning the $1\sigma$ to $2\sigma$ GRAVITY confidence level for $n_0^{\rm max}$, while the dotted lines use a more optimistic normalisation based on mass segregation, Eq.~\eqref{eq:n0_ceiling}.  The black line is the LISA power-law-integrated sensitivity~\cite{Thrane2013} ($T_{\rm obs}=4\,$yr, ${\rm SNR}=5$), built from the instrument noise plus the Galactic white-dwarf confusion foreground~\cite{Robson2019}.}
    \label{fig:OmegaGWemrbs}
\end{figure}

Armed with the MBH mass function from Sec.~\ref{sec:scaling}, the EMRBs rate and their energy spectrum, we can plug everything in Eq.~\eqref{eq:omega_general} and find the stochastic gravitational wave background due to EMRBs. Since the inferred mass function is local, using it at finite redshift amounts to neglecting redshift evolution of the MBH mass function. We thus conservatively limit our redshift integrals to $z \leq 3$, for which we consider this a plausible assumption.\footnote{Extending to higher redshifts using a red-shift independent MBH mass function does not significantly change our results, indicating that for most plausible redshift evolutions this covers the bulk of the signal contribution.}
In Fig.~\ref{fig:OmegaGWemrbs} we show our results for a population of $m=10 \, M_\odot$, adopting the upper limits from GRAVITY for the sBH BW profile rescaled to other galaxies according to Eq.~\eqref{eq:n0_prescription} (solid) and using the phenomenological upper limit Eq.~\eqref{eq:n0_ceiling}.
To assess the detectability of the EMRB background we compare it with the LISA power law sensitivity (PLS, solid black line)~\cite{Thrane2013}, the details of which are given in App.~\ref{app:LISA-PLS}. We conclude that this SGWB of extraglactic EMRBs is relevant for LISA only under the most optimistic assumptions for both the MBH mass function and the sBH population (solid blue). 

Our fiducial upper limit (solid blue) are very similar (up to order $\mathcal{O}(1)$ factors in terms of strain) to the EMRB background estimated in Ref.~\cite{Toonen:2009qi}, which used a sBH profile calibrated to the galactic center model of \cite{Alexander:2008tq} and the MBH mass function of Ref.~\cite{Aller:2002rp}. The results in Ref.~\cite{Oliver:2025irg} seem to fall in a similar ballpark, depending on the modelling assumptions for the burst signal. An even larger background is found in \cite{OLeary:2008myb}, due to an increased mass fraction of sBHS. This illustrates that, despite some remaining uncertainties, the astrometric data from the center of our galaxy is providing informative and complementary constraints for LISA science.

Furthermore, it is interesting to notice that under the self-similar scaling the EMRB background is remarkably flat in
frequency. Proposed $\mu$Hz missions such as
$\mu$Ares~\cite{Sesana:2019vho} or
ASTROD-GW\cite{Ni:2012eh} could therefore gain access to the bulk of this population.

\subsection{EMRI}

We now pass to the case of EMRIs. In this case, the expected rate per galaxy is much smaller than for EMRBs, because the physical requirement for producing an EMRI is much more restrictive. An EMRB only requires an object to be scattered onto an orbit with a sufficiently small periapsis so that it emits a detectable burst of gravitational radiation during a close passage near the massive black hole. By contrast, an EMRI requires that the compact object not only reach the strong-field region, but also remain bound and inspiral gradually under gravitational-wave emission before two-body relaxation can scatter it back out of the loss cone or drive it into a direct plunge. Most objects that enter the burst-producing region do not become EMRIs, because they either experience only one or a few burst-like passages, and then they are scattered away before gravitational radiation dominates their evolution, or they plunge without completing a long inspiral. For this reason, the EMRI rate per galaxy is naturally expected to be several orders of magnitude below the EMRB burst rate, even when both are sourced by the same underlying population of compact objects in the nuclear stellar cluster.

\paragraph{Event rate per galaxy.}

\begin{figure}
    \centering
    \includegraphics[width=\linewidth]{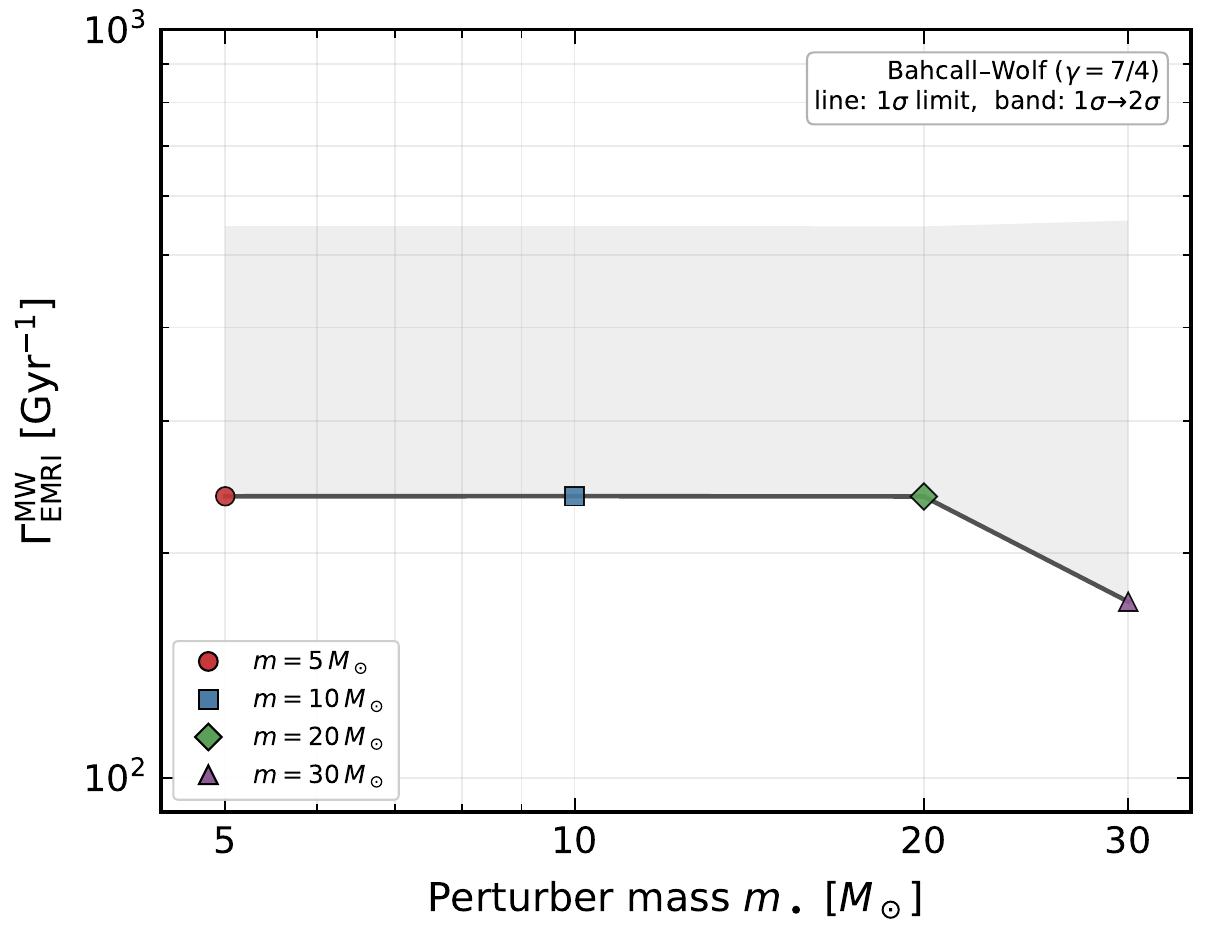}
    \caption{S2-derived upper limit on the per-galaxy EMRI rate for a Milky-Way-like MBH ($M_\bullet = 4\times10^{6}M_\odot$). The stellar-mass black holes are assumed to form a single self-consistent BW cusp that supplies both the EMRI sources and the relaxation scatterers ($\mu = m_{\rm sc} = m$), normalised to the strongest GRAVITY/S2 limit on $n_0$. The solid curve uses the $1\sigma$ limit; the band spans $1\sigma\to 2\sigma$.}
    \label{fig:EMRI_galaxy}
\end{figure}

Fig.~\ref{fig:EMRI_galaxy} shows the S2-derived upper limit on the per-galaxy EMRI rate for a Milky-Way-like massive black hole ($M_\bullet = 4\times10^{6}~M_\odot$), using Eq.~\eqref{eq:emri_rate_per_galaxy}, as a function of the perturber mass $m_\bullet$, assuming a BW profile. The $1\sigma$ bound is remarkably flat across perturber mass, remaining near $238~\mathrm{Gyr}^{-1}$ from $5$ to $20\,M_\odot$ before dropping slightly at $30\,M_\odot$. This near-independence follows from a near-exact cancellation between the EMRI rate and the binding S2 constraints. The mild downturn at $30 M_\odot$ marks where inclination, rather than precession, becomes the limiting S2 observable.

Our upper bounds are in the same ballpark as previous rate estimates in the literature, notably the rate estimates of 1 - 100/Gyr quoted in the LISA review paper~\cite{Babak:2017tow}, indicating that current astrometric (non-)detections remain fully compatible with standard steady-state EMRI predictions while beginning to probe the most interesting parameter space. For example, Ref.~\cite{AmaroSeoane:2011zz}, run a significant number of direct-summation N-body simulations, calibrating a much faster orbit-averaged Fokker–Planck code, and found ~250 events per Gyr per Milky Way type galaxy, using the stellar mass normalization for the inner parsec of the Galactic center at the time. Our results are also fully consistent with \cite{Rom:2024nso}. The $1\sigma$ GRAVITY bounds translate to $f_\bullet \sim 10^{-2}$, for which Ref.~\cite{Rom:2024nso} find a galactic rate of 260/Gyr. On the contrary, our upper limits are in some tension with the prediction of about $10^4$ events/Gyr in Ref.~\cite{Emami:2019uty}, indicating that the galactic sBH model~\cite{Emami:2019mzi} may be in some tension with the GRAVITY limits.

\begin{figure}
    \centering
    \includegraphics[width=\linewidth]{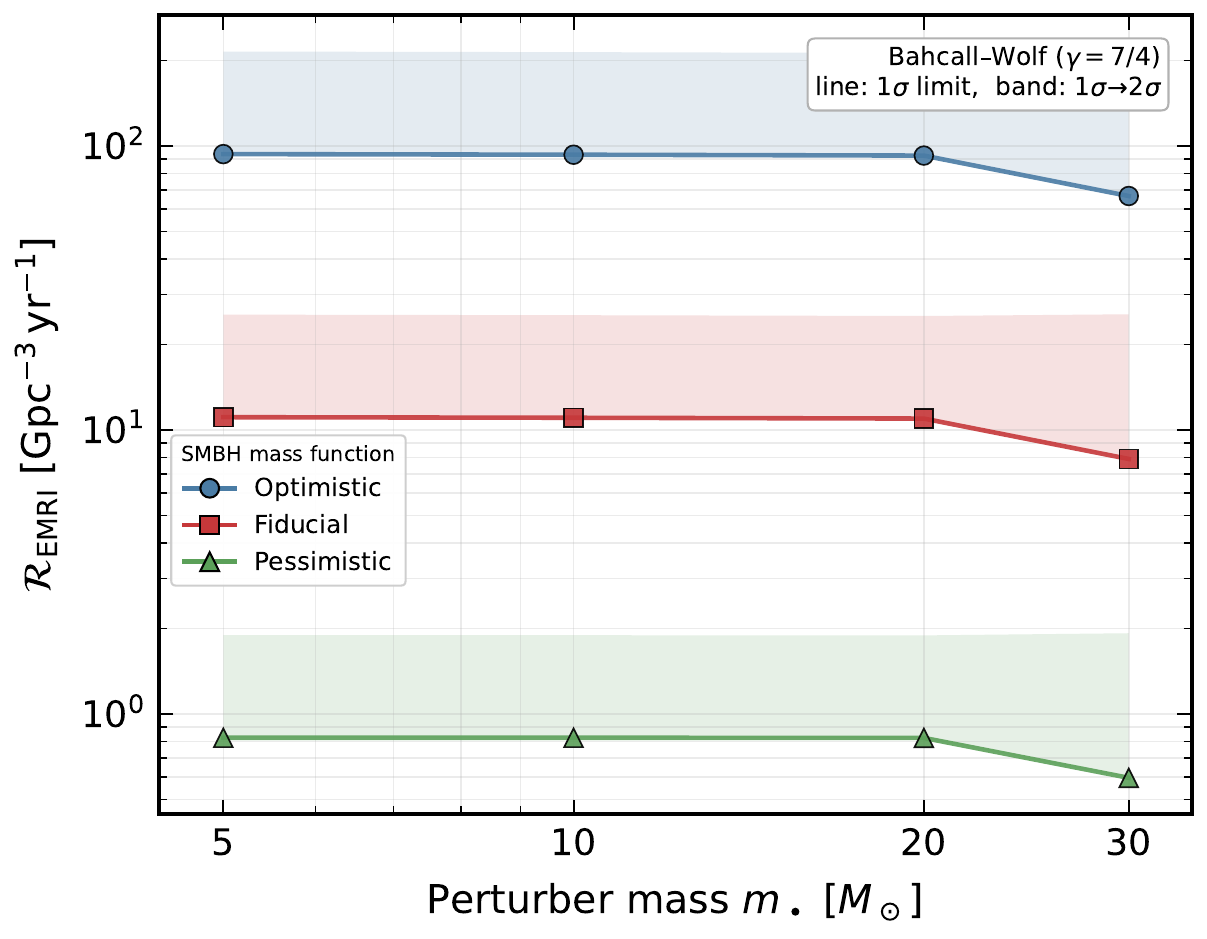}
    \caption{S2-derived upper limit on the volumetric (extra-galactic) EMRI rate density $\mathcal{R}{\rm EMRI}$, versus perturber mass $m$. Each per-galaxy BH-scattering rate (Fig.~\ref{fig:EMRI_galaxy}) is convolved with an MBH mass function — optimistic, fiducial, or pessimistic — over $M\bullet = 10^{4}$–$10^{9}M_\odot$. As in Fig.~\ref{fig:EMRI_galaxy}, the stellar-mass black holes form a self-consistent BW normalised to the strongest GRAVITY/S2 limit on $n_0$.}
    \label{fig:EMRI_extragalaxy}
\end{figure}

\paragraph{Extragalactic event rate.} 
Once we have computed the rate per galaxy, we can convolve it with cosmology and MBH mass functions, to obtain the extragalactic EMRI rate. We start with the rate per cosmic volume, $d\dot N_\text{EMRI}/dV_c$, see Eq.~\eqref{eq:observed_total_event_rate}. Varying the MBH mass function as described in Sec.~\ref{sec:scaling} and using the self-similar scaling for the sBH density in Eq.~\eqref{eq:n0_prescription}, we obtain upper limits on the EMRI formation rate informed by the GRAVITY anchor shown in  Fig.~\ref{fig:EMRI_extragalaxy}. The order-of-magnitude spread reflects the uncertain low-mass end of the MBH population, which dominates the integral, rather than the peak of the mass distribution.
As for the galactic rate, the bound is nearly independent of the perturber mass.

\begin{figure}
    \centering
    \includegraphics[width=\linewidth]{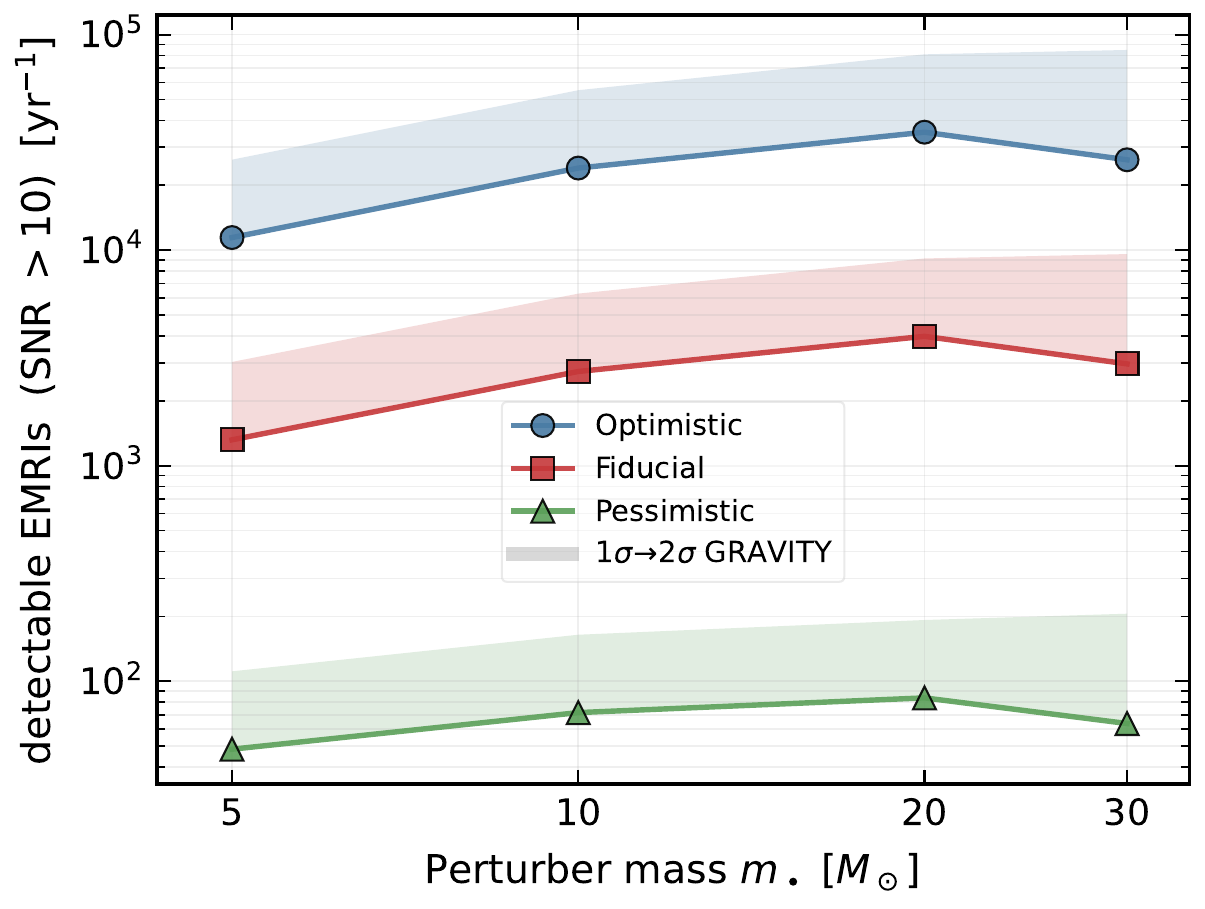}
    \caption{S2-derived upper limit on rate of EMRI detectable by LISA at $\text{SNR} > 10$, versus perturber mass $m$. Each per-galaxy BH-scattering rate (Fig.~\ref{fig:EMRI_galaxy}) is convolved with an MBH mass function — optimistic, fiducial, or pessimistic — over $M\bullet = 10^{4}$–$10^{9}M_\odot$. As in Fig.~\ref{fig:EMRI_galaxy}, the stellar-mass black holes form a self-consistent BW normalised to the strongest GRAVITY/S2 limit on $n_0$; the solid curves use the $1\sigma$ limit and the shaded bands span $1\sigma\to2\sigma$.}
    \label{fig:EMRI-nr-LISA}
\end{figure}

To determine the fraction of these events that are detectable by LISA, we impose and SNR threshold of 10.
We assume that the EMRI population has reached a steady state equilibrium, in which the rate of EMRI formation is equal to the rate of mergers.
We expect the EMRIs detectable by LISA to be already relatively circularized, so that we approximate the characteristic strain by the usual inspiral expression~\cite{Maggiore:2007ulw},
\begin{align}
 h_c^\text{EMRI}(f) = \frac{8 (\pi f)^{2/3}}{\sqrt{5} \, D_L(z) c^4}  ( (1 + z) G {\cal M}_c)^{5/3}  N_\text{cyc}^{1/2} \,,
\end{align}
where ${\cal M}_c = m_\bullet^{3/5} M_\bullet^{2/5}$ is the  chirp mass, $D_L(z)$ the luminosity distance and the number of observed cycles at frequency $f$ is given by $N_\text{cyc} = f\times\text{min}(T_\text{obs}, f/\dot f)$. Contrasting this with the LISA noise curve $h_c^n$, see App.~\ref{app:LISA-PLS}, gives the SNR as~\cite{Maggiore:2007ulw}
\begin{align}
 \text{SNR}^2 = \frac{16}{5} \int_{f_\text{low}}^{f_\text{isco}} d\ln f \, \left( \frac{h_c^\text{EMRI}}{h_c^n} \right)^2 \,,
 \label{eq:SNR-emri}
\end{align}
with the integration range running from the lowest frequencies detectable by LISA, $f_\text{low} \sim 10^{-6}$~Hz to the largest frequency generated by the EMRI at the inner most circular orbit (isco), $f_\text{isco} = (1 + z)^{-1} c^3/(\pi 6^{3/2} G M_\bullet) $. We additionally ensure, by modifying the lower integration boundary, that the SNR is computed over maxially 4 years of observation time.
Setting an SNR threshold, Eq.~\eqref{eq:SNR-emri} translates to a redshift reach $z_\text{max}$ for fixed sBH and MBH masses. We can now perform the volume integral in Eq.~\eqref{eq:observed_total_event_rate} up to this redshift, to obtain an upper limit on the number of detectable events in LISA, shown in Fig.~\ref{fig:EMRI-nr-LISA}.

Our upper limits  sit in the upper range  of published EMRI detection forecasts, which span from essentially zero up to a few thousand events per year
\cite{Babak:2017tow,Bonetti:2020jku}. A like-for-like comparison with Ref.~\cite{Rom:2024nso}
is also instructive, as they predict $\simeq 2\times10^{3}$ detectable EMRIs over a $4$-yr mission at
${\rm SNR}>20$ ($\simeq 10^{4}$ at ${\rm SNR}>8$). Evaluating our pipeline at the same
threshold and with the same MBH mass function (the fit of Ref.~\cite{Babak:2017tow} to the
Barausse~2012 model), we obtain $\simeq 1.2\times10^{4}$ ($3.6\times10^{4}$ at ${\rm SNR}>8$),
a factor of a few higher. Notably, the two calculations share essentially the same EMRI-producing cusp density --- the GRAVITY $1\sigma$ bound coincides with the universal, $f_\bullet$-independent branch of Ref.~\cite{Rom:2024nso} to within a factor $\lesssim 2$ --- so that the residual difference should reflect the self-confusion suppression included in their forecast rather than a difference in the underlying
remnant abundance.

\paragraph{SGWB.}
Given the total event rate Eq.~\eqref{eq:observed_total_event_rate}, we can convolute it with the emitted energy spectrum per event to compute  the stochastic GW background via Eq.~\eqref{eq:omega_general}. For the energy spectrum, we rely again on the circular, quadrupole-driven Newtonian inspiral~\cite{Maggiore:2007ulw},
\begin{align}
 \frac{dE}{df_s} = \frac{(\pi G)^{2/3}}{3}\,\mathcal{M}_c^{5/3}\,f_s^{-1/3} \,.
\end{align}
The $z$-integral in Eq.~\eqref{eq:omega_general} can be performed analytically. The upper integration boundary is set by the maximal redshift (for fixed $m_\bullet$ and $M_\bullet$) indicating when the maximal signal frequency is redshifted below the LISA band, $f_\text{isco} < f_\text{low}$ and requiring $z \leq 3$. To subtract resolvable events, we additionally impose a minimal redshift $z_\text{min} = z(\text{SNR} = 20)$. We note that our prescription of computing the SNR, see Eq.~\eqref{eq:SNR-emri}, assumes that the merger falls within the observation time, which is not necessarily the case. Our subtraction scheme may thus lead to slight underestimation of the unresolved SGWB.

\begin{figure}
    \centering
    \includegraphics[width=\linewidth]{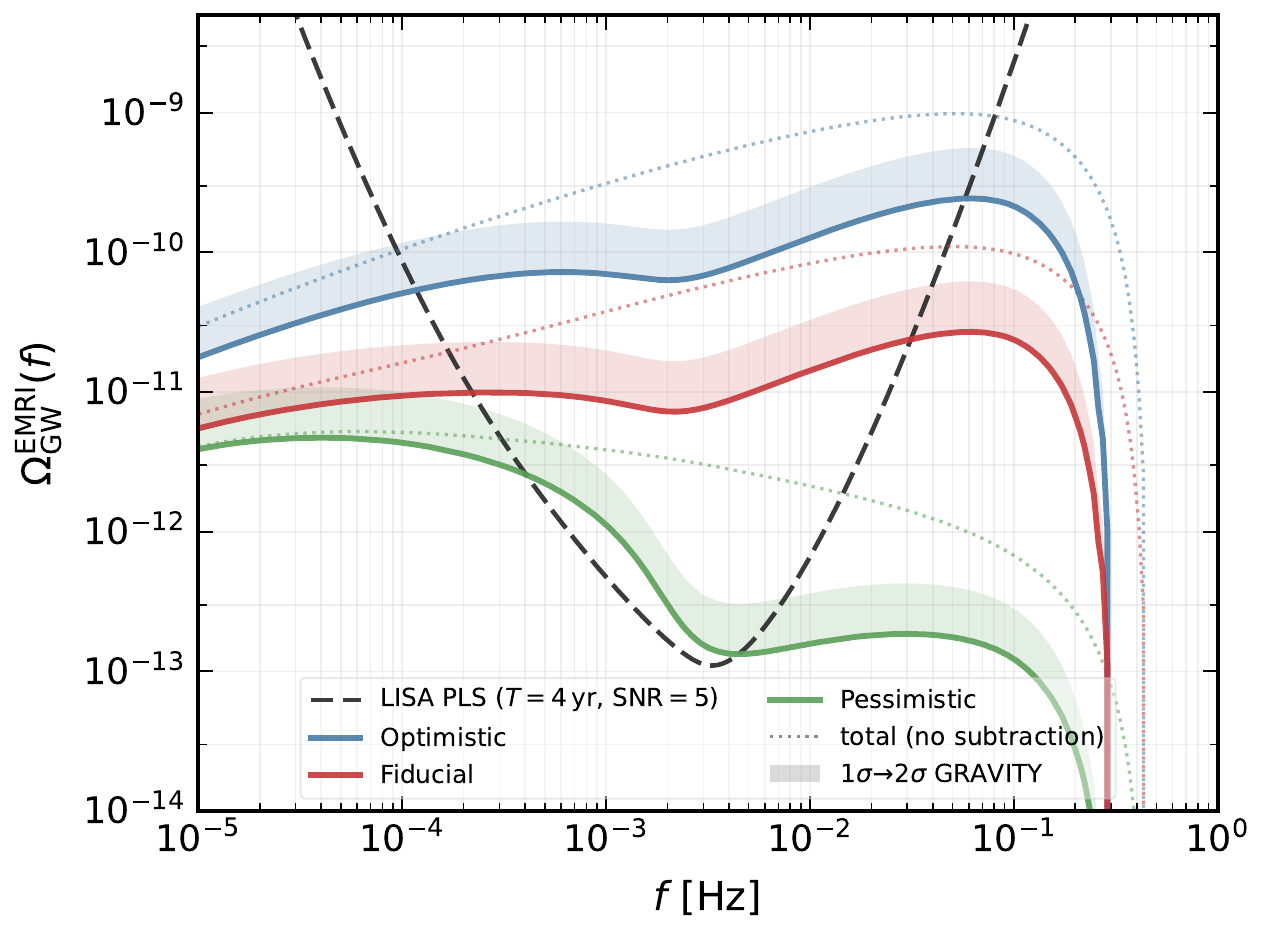}
    \caption{Upper limit on the extragalactic EMRI stochastic gravitational-wave background, $\Omega_{\rm GW}^{\rm EMRI}(f)$, informed by the S2/GRAVITY constraint on the stellar-mass black-hole number density, compared with the LISA power-law-integrated sensitivity (PLS; $T = 4$ yr, ${\rm SNR}_{\rm thr} = 5$, black dashed). The three coloured curves show the \emph{unresolved} residual background, with the shaded bands spanning the $1\sigma$ to $2\sigma$ GRAVITY confidence level for $n_0^{\rm max}$. Individually detectable EMRIs, which we define as those whose sky- and orientation-averaged single-source signal-to-noise ratio exceeds ${\rm SNR} = 20$ over the $4$-yr mission, are subtracted (dotted lines show the total, unsubtracted background for reference). Each curve assumes a single Bahcall--Wolf cusp normalised to the strongest GRAVITY/S2 limit on $n_0$ and rescaled self-similarly via Eq.~\ref{eq:n0_prescription}, with stellar-mass black holes ($m_\bullet = 10\,M_\odot$) acting as both EMRI sources and scatterers. The curves differ only in the assumed MBH mass function---optimistic, fiducial, or pessimistic---integrated over $M_\bullet = 10^{4}$--$10^{9}\,M_\odot$.}
    \label{fig:omegaEMRI}
\end{figure}

Our GRAVITY informed upper limits on the SGWB from unresolved EMRIs are shown in  Fig.~\ref{fig:omegaEMRI} for $m_\bullet = 10~M_\odot$.  Even at the S2-limited normalisation the signal lies well above the LISA power-law-integrated sensitivity ($T = 4~{\rm yr}$, ${\rm SNR}_{\rm thr} = 5$) for the optimistic and fiducial mass functions, and marginally so for the pessimistic one. S2 astrometry therefore can accommodate a LISA-detectable EMRI background, and the most important remaining uncertainties are the MBH mass function and the sBH mass profile, though the depletion factor matters less than in the EMRB case since we are less sensitive to the lower integration boundary in $r_p$.

The comparison with the PLS curve, as shown in Fig.~\ref{fig:omegaEMRI}, is the relevant figure of merit to determine the detectability of this SGWB, which would yield key insights on black hole population models. It is also key to determin if this SGWB my hide other, cosmological sources of stochastic backgrounds. On the other hand, to determine if this SGWB will obstruct measurements of transient signals, the corresponding characteristic strain
\begin{align}
 h_c = \frac{H_0}{\sqrt{2} \pi f} (3 \Omega_\text{gw})^{1/2} \,,
\end{align}
should be contrasted directly with the LISA instrumental noise curve. From this we expect at most a marginal impact on the ability of LISA to measure transient signals.

Our upper limits on the EMRI SGWB fall broadly within the ball park of earlier estimates~\cite{Barack:2004wc,Bonetti:2020jku,Rom:2024nso}. The range of these predictions, similar to the variation in our limits, can be traced back to two main uncertainties: the MBH mass function at relatively small MBH masses and the subtraction scheme employed to determine the unresolvable SGWB. In particular, the result in \cite{Rom:2024nso}, $h_c\simeq2.1\times10^{-21}(f/2.5\,{\rm mHz})^{-1.26}$, is a confusion-noise level obtained from source counts. This method retains the full spectrum at low frequencies $f \lesssim$~mHz, where the frequency bins are crowded and sources overlap (the dotted curves in Fig.~\ref{fig:omegaEMRI}), but suppresses power at higher frequencies, where the bins become sparse and individual sources can be resolved. On the other hand, our subtraction scheme removes the EMRI emission at all frequencies for every host within the resolvability horizon, including the early, slowly evolving inspiral stages whose signal-to-noise ratio is individually below threshold, whereas a source-by-source removal, as in Ref.~\cite{Rom:2024nso}, retains this sub-threshold emission in the background. Our subtraction scheme therefore removes somewhat more power in the central sensitivity regime of LISA. Overall, due to the combined effect of the mass function employed and the subtraction criterion, our 1$\sigma$ limit falls a factor below the prediction of \cite{Rom:2024nso}.

The arguably most detailed EMRI background computation to date is that of
Ref.~\cite{Bonetti:2020jku}, who build on the population-synthesis catalogues of
Ref.~\cite{Babak:2017tow}, reconstruct the full inspiralling population via the
continuity equation, sum the eccentric harmonics of kludge waveforms, and remove
individually resolvable sources with ${\rm SNR}>20$, the same threshold adopted in this work.
Despite the entirely different anchoring, population-synthesis scenarios there versus a
GRAVITY-calibrated upper limit here, the residual backgrounds agree remarkably well. The slighly different behaviour at low frequencies can be traced back to the finite backward-integration window (their population only includes inspiral stages within a few
centuries of plunge, whereas the Eq.~\eqref{eq:omega_general} accounts for all stages) and to the
eccentric harmonics concentrating the early-inspiral power near the periapsis frequency. Interestingly, the fact that our simplified circular-inspiral background is close to the results of Ref.~\cite{Bonetti:2020jku},
shows that the approximation is adequate for the amplitude in band. The signal is dominated by the least eccentric inspirals, around the more massive black holes, which reach the band closer to circularisation, while the most eccentric sources radiate at higher harmonics and contribute little where LISA is most sensitive.

\section{Conclusions}
\label{sec:conclusions}

The recent astrometric data obtained by precision measurements of the center of our galaxy by the GRAVITY observatory provides key insights on the sBH distribution in the inner ${\cal O}(1 - 100)$~mpc around the central black hole Sgr A$^*$. This is precisely the region of relevance for sourcing detectable GW burst signals from fly-by orbits, i.e.\ EMRBs, and moreover coincides with the formation region of extreme mass ratio inspirals (EMRIs). Both events yield a sizeable GW signal in the mHz regime, and are key targets for the LISA mission.

In this paper we apply precession and inclination measurements of the S2 orbit by GRAVITY to obtain upper limits on the sBH profile, which immediately translate to upper limits on the galactic EMRB rate of less than one per year (see Fig.~\ref{fig:EMRBrates}), which is on the lower end of earlier expectations. The galactic EMRI rate is many orders of magnitude smaller.

Turning to extragalactic sources introduces systematic uncertainties due to extrapolation of the sBH profile to other galaxies, as well as due to large uncertainties in the MBH mass function, in particular at lower MBH masses which are of particular importance to the cosmological EMRB and EMRI signals. We take this into account by presenting a fiducial self-similar scaling of the sBH profile and a fiducial MBH mass function, supplemented by very generous alternative scalings to track this uncertainty.

Our results in Figs.~\ref{fig:OmegaGWemrbs}, \ref{fig:EMRI-nr-LISA}, \ref{fig:omegaEMRI} show upper limits on the number of detectable EMRI events in LISA as well as upper limits on the expected gravitational wave background from both EMRBs and EMRIs.  In all cases, our limits fall within the ballpark of previous estimates for these quantities, demonstrating the transformative impact that galactic astrometric data can have on these predictions. We provide analytical expressions, in particular for the scaling with $n_0$, $\Gamma_{\rm EMRI}\propto n_0^{6/5}$ and $\Gamma_{\rm EMRB}\propto n_0^{1.6}$, enabling a transparent re-scaling of our results in the advent of stronger astrometric limits.

The near-term astrometric outlook is particularly promising. The 2026 apocentre passage of S2 offers enhanced sensitivity to stochastic perturbations from individual stellar-mass black holes
\cite{Bordoni:2025mli}. In addition, the recent discovery of S301, on an
$8.7$-yr orbit with a pericentre approximately ten times smaller than that of S2, introduces a complementary probe of the gravitational potential much closer to Sgr~A$^*$ \cite{Dayem:2026ktt}. Continued GRAVITY observations,
together with future spectroscopy, may enable joint analyses of S2, S301,
and additional faint S-stars. Such measurements could help separate smooth
extended mass, stochastic compact-object perturbations, and relativistic
precession across different radial scales. A detection of granularity would replace the present one-sided constraint on the local cusp normalisation with an empirical measurement, substantially narrowing the allowed Galactic EMRI and EMRB rates. Conversely, detections or
upper limits from LISA would probe the inward continuation and depletion of
the cusp and, for cosmological sources, its scaling across the MBH
population.

Our work is complementary to constraints on the massive black-hole population from pulsar-timing arrays~\cite{EPTA:2023xxk,NANOGrav:2023hfp}, see e.g.~\cite{Sato-Polito:2023gym,Goncharov:2024htb,Sesana:2025udx} for recent discussions on the compatibility of the observed GW signal with the expected MBH population.
The nano-Hz stochastic background traces the demographics of the most massive black holes ($M_\bullet\gtrsim10^{8}\,M_\odot$) through their binary mergers, whereas the S2/GRAVITY calibration developed here, propagated to the LISA band, is most sensitive to the low-mass end ($M_\bullet\lesssim10^{5}\,M_\odot$). The two probes therefore bracket the central black-hole mass function
from opposite ends. A joint use of nano-Hz and milli-Hz backgrounds,
anchored to the same local demographics, thus offers a route to testing black-hole demographics
across the full mass spectrum.  This complementarity establishes stellar
astrometry and millihertz gravitational waves as two views of the same compact-remnant population in galactic nuclei.

\vspace{.2cm}
\textit{Acknowledgments.} We thank Boaz Katz and Kfir Blum for discussions during the early stages of this project. We also thank Matteo Bonetti for useful comments on the draft. AC is supported by an ERC STG grant (``AstroDarkLS'', grant No. 101117510). AC acknowledges the Weizmann Institute of Science for hospitality at different stages of this project and the support from the Benoziyo Endowment Fund for the Advancement of Science. Both authors acknowledge the support by the European Research Area (ERA) via the UNDARK project (project number 101159929).

\appendix

\section{Technical Details about our simulations and constraints from S2 motion}
\label{app:simulations}

The S2 star is evolved under the total acceleration
  \begin{equation}
      \ddot{\vec{r}} = \vec{a}_{\rm Newt} + \vec{a}_{\rm 1PN} + \vec{a}_{\rm pert},
  \end{equation}
  where the Newtonian term is $\vec{a}_{\rm Newt}=-GM_\bullet\vec{r}/r^3$, the
  1PN Schwarzschild correction is
  \begin{equation}
      \vec{a}_{\rm 1PN} = \frac{GM_\bullet}{r^2 c^2}\left[\left(\frac{4GM_\bullet}{r}-v^2\right)\hat{r} + 4(\hat{r}\cdot\vec{v})\vec{v}\right],
  \end{equation}
and the perturber contribution is
\begin{equation}
      \vec{a}_{\rm pert} = -\sum_{i=1}^{N_p} \frac{Gm_i(\vec{r}-\vec{r}_i)}{(|\vec{r}-\vec{r}_i|^2+\epsilon^2)^{3/2}},
\end{equation}
with softening length $\epsilon=0.007$~AU to avoid numerical divergences. Each perturber is assigned a Keplerian orbit around Sgr~A* with
  \begin{itemize}
      \item Semi-major axis sampled from $n(a)\propto a^{-\gamma}$ via inverse transform;
      \item Eccentricity from thermal distribution: $e=\sqrt{u}$, $u\sim\mathcal{U}(0,1)$;
      \item Isotropic orientation: $\cos i\sim\mathcal{U}(-1,1)$, $\Omega,\omega,M_0\sim\mathcal{U}(0,2\pi)$.
  \end{itemize}
Perturber positions are evolved analytically using Kepler's equation (no perturber-perturber interactions), and pre-tabulated for efficiency. We use a velocity Verlet integrator with predictor-corrector iterations to handle the velocity-dependent 1PN force. The timestep is $\Delta t = P_{\rm S2}/50000$, giving $\sim 10^5$ steps over two orbits. Energy conservation is verified to $\Delta E/E < 10^{-8}$ for the unperturbed case. At each pericenter passage, we compute the eccentricity vector $\vec{e}=(\vec{v}\times\vec{h})/\mu - \hat{r}$ and angular momentum $\vec{h}=\vec{r}\times\vec{v}$. The in-plane precession $\Delta\omega$ is measured as the angle between successive eccentricity vectors projected onto the orbital plane; the inclination change $\Delta i$ is the angle between successive $\vec{h}$ vectors.

With no perturbers, the code recovers the theoretical Schwarzschild precession $\Delta\omega_{\rm GR}=6\pi GM_\bullet/[c^2 a(1-e^2)]=12.26$~arcmin/orbit to within 0.3\%. The baseline integration shows zero inclination drift ($\Delta i < 10^{-6}$~arcsec).

\section{Derivation of upper limits on the density normalisation}
\label{app:n0_method}

This appendix describes the procedure used to derive upper limits on the
stellar-mass black-hole cusp density normalisation~$n_0$ from GRAVITY
observations of the S2 orbit, for each combination of cusp slope~$\gamma$ and
individual black-hole mass~$m$.

\subsection{Simulation sweep}
\label{app:sweep}

For a given pair $(\gamma,\,m)$ we run the $N$-body integrator
(\texttt{s2\_orbit}) at a set of discrete perturber numbers
$\{N_p^{(k)}\}_{k=1}^{K}$, with $K = 7$ values chosen to bracket the expected
$N_\mathrm{max}$ for both the precession and inclination constraints.  Because
heavier perturbers produce larger perturbations per object, the sweep range is
shifted downward with increasing~$m$:

\begin{table}[h]
\centering
\begin{tabular}{c l}
\toprule
$m\;[M_\odot]$ & $N_p$ sweep values \\
\midrule
5  & 50,\; 100,\; 200,\; 400,\; 600,\; 800,\; 1000 \\
10 & 25,\; 50,\; 100,\; 200,\; 300,\; 400,\; 500 \\
20 & 10,\; 25,\; 50,\; 100,\; 150,\; 200,\; 300 \\
30 & 5,\;\; 10,\;\; 25,\;\; 50,\;\; 75,\;\; 100,\; 150 \\
\bottomrule
\end{tabular}
\caption{Sweep ranges used in the parameter scan.  Each simulation is run with
$N_\mathrm{MC} = 40$ Monte Carlo realisations of the perturber population,
drawing semi-major axes from $n(a)\propto a^{-\gamma}$ in the range
$60$--$3900\,\text{AU}$, thermal eccentricities $f(e)=2e$, and isotropic
orientations.}
\label{tab:sweep_ranges}
\end{table}

\noindent
At each sweep point we record the mean absolute precession shift
$|\Delta\omega|$ and mean inclination change $\Delta i$ (with their standard
deviations across realisations), providing $K$ data points for each observable
as a function of $N_p$.

\subsection{Scaling relations and the coefficients \texorpdfstring{$A$}{A} and
\texorpdfstring{$B$}{B}}
\label{app:scaling}

The two observables follow distinct scaling laws with the number of perturbers,
rooted in different physical mechanisms.

\paragraph{Precession: coherent regime.}
Each perturber contributes a small, approximately systematic apsidal shift to
S2's orbit.  These contributions add up \emph{coherently}, so the net
precession perturbation scales linearly:
\begin{equation}
\boxed{|\Delta\omega| \;=\; A \times N_p}\,,
\label{eq:precession_scaling}
\end{equation}
where $A$ is the mean precession shift per perturber per orbit, in arcmin.
Physically $A \propto m$, since each perturber's pull scales with its mass; the
coefficient also depends on~$\gamma$ through the radial distribution of
perturbers relative to S2's orbit.

\paragraph{Inclination: random-walk regime.}
Each perturber exerts a gravitational torque on S2's orbital plane, tilting it
by a small random angle.  Because the perturbers have random, isotropic
orientations, successive kicks are uncorrelated and the root-mean-square
inclination change grows as a random walk:
\begin{equation}
\boxed{\Delta i \;=\; B \times \sqrt{N_p}}\,,
\label{eq:inclination_scaling}
\end{equation}
where $B$ is the RMS inclination kick per $\sqrt{\text{perturber}}$ per orbit,
in arcsec.  Like~$A$, the coefficient $B \propto m$ and depends on~$\gamma$.

\begin{figure*}[t]
  \centering
  \includegraphics[width=2\columnwidth]{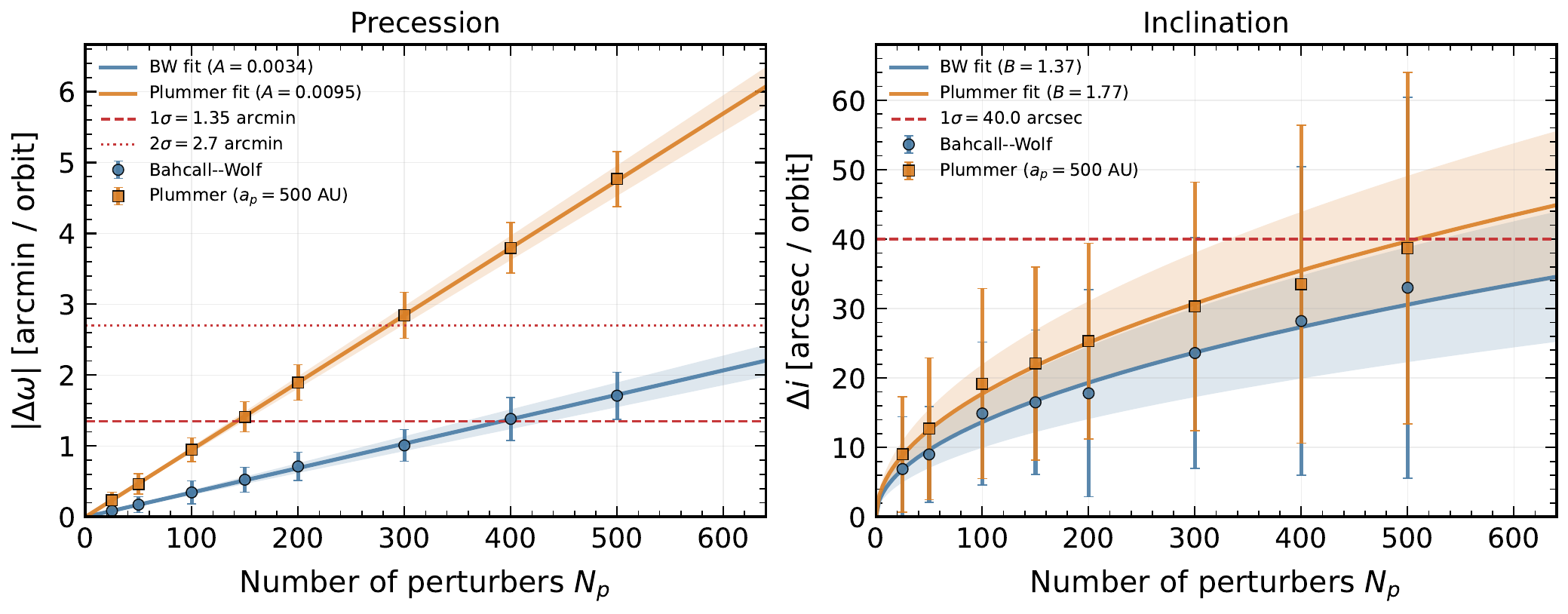}
  \caption{S2 orbital perturbations from a stellar-mass black-hole population
with individual mass $m = 10\,M_\odot$.  \textit{Left:} absolute precession
shift $|\Delta\omega|$ (arcmin per orbit) versus the number of perturbers~$N_p$,
for a Bahcall--Wolf cusp (blue circles) and a Plummer sphere ($a_p = 500\,$AU;
orange squares).  Each point is the mean over 50 Monte Carlo realisations, with
error bars showing the standard deviation; solid lines are weighted
least-squares fits to $|\Delta\omega| = A\,N_p$, with shaded $1\sigma$ bands on
$A$.  The Plummer profile yields $A = 0.0095\,$arcmin, $\simeq 2.8\times$ larger
than the Bahcall--Wolf value $A = 0.0034\,$arcmin, because it concentrates more
mass within S2's orbit.  Horizontal red lines mark the GRAVITY $1\sigma$ and
$2\sigma$ precession thresholds ($\sigma_{\Delta\omega} = 1.35$ and
$2.70\,$arcmin, from $f_\mathrm{SP} = 1.135 \pm 0.110$ \cite{GRAVITY2024}).
\textit{Right:} inclination change $\Delta i$ (arcsec per orbit), with fits to
$\Delta i = B\,\sqrt{N_p}$, giving $B = 1.37\,$arcsec (Bahcall--Wolf) and
$B = 1.77\,$arcsec (Plummer); red lines mark the $1\sigma$ and $2\sigma$
inclination thresholds ($\sigma_{\Delta i} = 40$ and $80\,$arcsec, from
GRAVITY's $\approx 30\,\mu$as astrometric accuracy \cite{GRAVITY2017} projected
to apocentre).  The intersection of each fit with a threshold defines
$N_\mathrm{max}$; for this reference case the precession channel is binding
($N_\mathrm{max}^{\rm prec}\simeq 400$ vs.\ $N_\mathrm{max}^{\rm incl}\simeq
850$ at $1\sigma$), while the granular inclination channel becomes binding only
for the heaviest perturbers ($m\gtrsim 20$--$30\,M_\odot$).}
\label{fig:s2_pert}
\end{figure*}

Both coefficients are determined by weighted least-squares fits through the
origin, using the Monte Carlo standard deviations as absolute errors:
\begin{align}
(A,\;\sigma_A) &= \operatorname{argmin}_A \sum_{k=1}^{K}
    \frac{\bigl(|\Delta\omega_k| - A\,N_p^{(k)}\bigr)^2}
         {\sigma_{|\Delta\omega_k|}^2}\,,
\label{eq:fit_A} \\[6pt]
(B,\;\sigma_B) &= \operatorname{argmin}_B \sum_{k=1}^{K}
    \frac{\bigl(\Delta i_k - B\,\sqrt{N_p^{(k)}}\bigr)^2}
         {\sigma_{\Delta i_k}^2}\,.
\label{eq:fit_B}
\end{align}
Figure~\ref{fig:s2_pert} illustrates these fits for the reference case
$m = 10\,M_\odot$, $\gamma = 7/4$, comparing the discrete Bahcall--Wolf cusp
with a smooth Plummer component.

\subsection{Derivation of \texorpdfstring{$N_\mathrm{max}$}{Nmax}}
\label{app:nmax}

The maximum number of perturbers compatible with GRAVITY is obtained by setting
the predicted perturbation equal to the measurement uncertainty. For the
apsidal advance we set $|\Delta\omega| = \sigma_{\Delta\omega}$ in
Eq.~\eqref{eq:precession_scaling},
\begin{equation}
N_\mathrm{max}^{(\mathrm{prec})} = \frac{\sigma_{\Delta\omega}}{A}\,,
\qquad
\frac{\delta N_\mathrm{max}}{N_\mathrm{max}} = \frac{\sigma_A}{A}\,,
\label{eq:Nmax_prec}
\end{equation}
with $\sigma_{\Delta\omega} = 1.35\,\text{arcmin}$ ($1\sigma$) or
$2.70\,\text{arcmin}$ ($2\sigma$), i.e.\ the width of the measured
Schwarzschild--precession factor $f_\mathrm{SP} = 1.135 \pm 0.110$
\cite{GRAVITY2024} times the relativistic precession
$\Delta\omega_{\rm GR}\simeq 12.3\,\text{arcmin}$ per orbit.  For the orbital
plane we set $\Delta i = \sigma_{\Delta i}$ in
Eq.~\eqref{eq:inclination_scaling},
\begin{equation}
N_\mathrm{max}^{(\mathrm{incl})} = \left(\frac{\sigma_{\Delta i}}{B}\right)^{\!2},
\qquad
\frac{\delta N_\mathrm{max}}{N_\mathrm{max}} = 2\,\frac{\sigma_B}{B}\,,
\label{eq:Nmax_incl}
\end{equation}
with $\sigma_{\Delta i} = 40\,\text{arcsec}$ ($1\sigma$) or
$80\,\text{arcsec}$ ($2\sigma$), the plane tilt that displaces S2 on the sky by
GRAVITY's astrometric accuracy of $\approx 30\,\mu$as \cite{GRAVITY2017,
Bordoni:2025mli} at apocentre, after the geometric projection
$\sin i\,|\sin(\nu+\omega)|\approx 0.66$.  Note the quadratic dependence
on~$B$: a $10\%$ uncertainty in~$B$ translates to a $20\%$ uncertainty in
$N_\mathrm{max}$.  For the reference case ($m = 10\,M_\odot$, $\gamma = 7/4$)
these give $N_\mathrm{max}^{(\mathrm{prec})}\simeq 400$ and
$N_\mathrm{max}^{(\mathrm{incl})}\simeq 850$ at $1\sigma$, so precession sets
the tighter (binding) bound.

\subsection{Conversion to density normalisation \texorpdfstring{$n_0$}{n0}}
\label{app:n0_conversion}

For the number-density profile the total number of objects in the radial shell
$[r_\mathrm{min},\, r_\mathrm{max}]$ is
\begin{equation}
N = \int_{r_\mathrm{min}}^{r_\mathrm{max}} n(r)\,4\pi r^2\,\mathrm{d}r
  = \frac{4\pi\,n_0\,r_0^{\gamma}}{3 - \gamma}
    \left(r_\mathrm{max}^{3-\gamma} - r_\mathrm{min}^{3-\gamma}\right),
\label{eq:N_from_n0}
\end{equation}
valid for $\gamma \neq 3$.  Inverting for~$n_0$,
\begin{equation}
\boxed{n_0 = \frac{(3-\gamma)\,N}{4\pi\,r_0^{\gamma}
    \left(r_\mathrm{max}^{3-\gamma} - r_\mathrm{min}^{3-\gamma}\right)}}\,,
\label{eq:n0_from_N}
\end{equation}
with $r_0 = 200\,\text{AU}$, $r_\mathrm{min} = 60\,\text{AU}$ and
$r_\mathrm{max} = 3900\,\text{AU}$.  Substituting $N = N_\mathrm{max}$ from the
precession or inclination constraint yields the corresponding upper limit
$n_0^\mathrm{max}$, and $\delta n_0$ follows by propagating $\delta
N_\mathrm{max}$ through the same linear relation.  At each $(\gamma,m)$ we adopt
the tighter (binding) of the two channels; for the reference case this gives
$n_0^\mathrm{max}\simeq 1.2\times10^{-7}\,\text{AU}^{-3}$ (precession $1\sigma$).

The two-panel plot (Fig.~\ref{fig:n0_s2}) displays $n_0^\mathrm{max}$ versus
$\gamma$ for each mass, with two distinct sources of uncertainty:
\begin{itemize}
\item \emph{Error bars} represent the statistical uncertainty on
$n_0^\mathrm{max}$ from the finite number of Monte Carlo realisations
($N_\mathrm{MC} = 40$ per sweep point): the scatter propagates into $\sigma_A$
and $\sigma_B$, and thence into $\delta N_\mathrm{max}$ via
Eqs.~\eqref{eq:Nmax_prec}--\eqref{eq:Nmax_incl}.  These are typically at the
few-percent level and quantify \emph{how precisely the simulation determines
the constraint}.
\item \emph{Shaded bands} span from the $1\sigma$ to the $2\sigma$ GRAVITY
measurement threshold: the solid curve uses the $1\sigma$ uncertainties
($\sigma_{\Delta\omega} = 1.35\,\text{arcmin}$,
$\sigma_{\Delta i} = 40\,\text{arcsec}$), while the band edge uses the $2\sigma$
values ($2.70\,\text{arcmin}$, $80\,\text{arcsec}$).  A larger allowed
perturbation admits more perturbers, yielding a weaker (higher) upper limit
on~$n_0$.
\end{itemize}
In summary, the error bars reflect the internal precision of our numerical
pipeline, while the bands reflect the external precision of the GRAVITY
observations.

\section{Isotropic distribution function of the sBH cusp}

\label{app:eddington}

The burst kernel of Eq.~\eqref{eq:emrb_rate_kernel} and the Monte-Carlo sampling of the S2 perturbers in Sec.~\ref{sec:gravity} both rest on the isotropic phase-space distribution function (DF)
$F_\bullet(\varepsilon)$ of a power-law cusp in the Keplerian potential of the central black hole. We derive it here by Eddington inversion, show that
it fixes the normalisation entering Eq.~\eqref{eq:emrb_rate_kernel}, and
that it reproduces the thermal eccentricities and $a^{-\gamma}$ semi-major axis distribution used in the S2 integration.

\paragraph{Setup.} The stellar-mass black holes follow the number-density profile $n_\bullet(a)=n_0 (a_0/a)^{\gamma}$ introduced in the main text, with mass density
$\rho_\bullet(a)=m_\bullet n_\bullet(a)$. At the radii of interest the supermassive black hole dominates the potential, so we work in the Keplerian
\emph{relative potential}
\begin{equation}
    \Psi(a)\equiv-\Phi(a)=\frac{G M_\bullet}{a},
    \label{eq:app_potential}
\end{equation}
in terms of which bound orbits carry a positive relative energy per unit
mass $\varepsilon=\Psi-\tfrac12 v^2>0$; for a Keplerian orbit
$\varepsilon=G M_\bullet/(2a)$, consistent with Eq.~\eqref{eq:emrb_rate_kernel}.
We assume an isotropic, steady-state DF that depends on phase space only
through $\varepsilon$.

\paragraph{Eddington inversion.}
For such a DF the Eddington formula relates $F_\bullet$ to the number density
expressed as a function of the potential~\cite{binney1987}
\begin{equation}
    F_\bullet(\varepsilon)=\frac{1}{\sqrt{8}\pi^{2}}
    \left[\int_{0}^{\varepsilon}
    \frac{d^{2}n_\bullet}{d\Psi^{2}}
    \frac{d\Psi}{\sqrt{\varepsilon-\Psi}}
    +\frac{1}{\sqrt{\varepsilon}}
    \left.\frac{dn_\bullet}{d\Psi}\right|_{\Psi=0}\right].
    \label{eq:app_eddington}
\end{equation}
Because the potential fixes $r=G M_\bullet/\Psi$, the density becomes a pure
power law in $\Psi$,
\begin{equation}
    n_\bullet(\Psi)=n_0 a_0^{\gamma}
    \left(\frac{\Psi}{G M_\bullet}\right)^{\!\gamma}
    \equiv N_0 \Psi^{\gamma},
    \,
    N_0\equiv\frac{n_0 a_0^{\gamma}}{(G M_\bullet)^{\gamma}} .
    \label{eq:app_nPsi}
\end{equation}
Here the constant $N_0$ is simply the coefficient obtained on rewriting
$(a_0/a)^{\gamma}$ as a power of $\Psi$. The density normalisation is still
$n_0=n_\bullet(a_0)$, while $a_0^{\gamma}/(G M_\bullet)^{\gamma}$ is the
Jacobian-like factor of the change of variable $r\to\Psi$, and carries the
\emph{central} mass $M_\bullet$. For $\gamma>1$ the boundary term in
Eq.~\eqref{eq:app_eddington} vanishes, since
$dn_\bullet/d\Psi\propto\Psi^{\gamma-1}\to0$ as $r\to\infty$. Substituting
$d^{2}n_\bullet/d\Psi^{2}=\gamma(\gamma-1)N_0\Psi^{\gamma-2}$ and rescaling
$\Psi=\varepsilon t$ reduces the integral to an Euler Beta function,
\begin{equation}
    \int_{0}^{1} t^{\gamma-2}(1-t)^{-1/2}\,dt
    =B\!\left(\gamma-1,\tfrac12\right)
    =\frac{\sqrt{\pi}\,\Gamma(\gamma-1)}{\Gamma(\gamma-\tfrac12)} .
    \label{eq:app_beta}
\end{equation}
Using $\gamma(\gamma-1)\Gamma(\gamma-1)=\Gamma(\gamma+1)$ we obtain the DF as
a pure power law in energy,
\begin{equation}
    F_\bullet(\varepsilon)=
    \frac{\Gamma(\gamma+1)}{2\sqrt{2}\pi^{3/2}\Gamma(\gamma-\tfrac12)}
    n_0 \frac{a_0^{\gamma}}{(G M_\bullet)^{\gamma}}
    \varepsilon^{\gamma-3/2}.
    \label{eq:app_FE}
\end{equation}
This is precisely the $F_\bullet(\varepsilon)$ that enters the burst kernel
Eq.~\eqref{eq:emrb_rate_kernel}. It is a power law of index $\gamma-3/2$
(e.g.\ $\varepsilon^{1/4}$ for the Bahcall--Wolf value $\gamma=7/4$),
non-negative for $\gamma\ge3/2$. Two features are worth stressing: the mass
appearing in $(G M_\bullet)^{\gamma}$ is the \emph{central} MBH, inherited
from the potential Eq.~\eqref{eq:app_potential}; and $F_\bullet$ is a
\emph{number} DF, so the individual remnant mass $m_\bullet$ does not appear
in it at all---it re-enters the EMRB calculation only through the detection
threshold $r_{p,\max}$ and the timescales $\tau_{\rm GW},\tau_J$ that set the
depletion weight $\mathcal W_{\rm B}$.

\paragraph{Velocities, eccentricities and semi-major axes.}
The same DF fixes the orbital distributions used to sample the S2 perturbers
in Sec.~\ref{sec:gravity}. At fixed radius the speed distribution follows
from $n_\bullet(r,v)\,dv=4\pi v^{2}F_\bullet(\Psi-\tfrac12 v^2)\,dv$,
\begin{equation}
    p(v\,|\,r)\propto v^{2}\left(1-\frac{v^{2}}{v_{\rm esc}^{2}}\right)^{\!\gamma-3/2},
    \, v_{\rm esc}=\sqrt{2G M_\bullet/r},
    \label{eq:app_speed}
\end{equation}
i.e.\ $(v/v_{\rm esc})^{2}\sim\mathrm{Beta}(3/2,\gamma-1/2)$, a shape
independent of radius. To obtain the distribution of orbital elements we use
the Keplerian density of states~\cite{binney1987},
$dN=F_\bullet(\varepsilon)(2\pi)^{2}T_r(\varepsilon,L)L\,dL\,d\varepsilon$,
where the radial period $T_r=2\pi a^{3/2}/\sqrt{G M_\bullet}$ depends on
$\varepsilon$ (equivalently $a$) alone---the defining property of the
Keplerian potential. At fixed energy $T_r$ factors out, leaving
$dN|_{\varepsilon}\propto L\,dL$; converting with
$L^{2}=G M_\bullet a(1-e^{2})$ gives the \emph{thermal} eccentricity
distribution
\begin{equation}
    p(e\,|\,\varepsilon)=2e,
    \label{eq:app_thermal}
\end{equation}
independent of $\gamma$. Integrating the density of states over $L$ at fixed
$a$, with $F_\bullet\propto\varepsilon^{\gamma-3/2}\propto a^{3/2-\gamma}$,
yields the semi-major axis distribution
\begin{equation}
    \frac{dN}{da}\propto a^{2-\gamma}.
    \label{eq:app_dNda}
\end{equation}
Equations~\eqref{eq:app_thermal}--\eqref{eq:app_dNda} are exactly the
sampling prescriptions adopted for the S2 integration---semi-major axes drawn
from $n(a)\propto a^{-\gamma}$ (a spatial power law times the shell volume
$a^{2}$) and thermal eccentricities $f(e)=2e$. The Eddington construction thus shows that this sampling is a
self-consistent phase-space realisation of the same cusp that sources the EMRB and EMRI signals.

\section{Analytical arguments for the expected power-law profiles of the sBH population}
\label{app:profiles}

The vertical lines in Fig.~\ref{fig:n0_s2} mark the two single-power-law slopes that we use below, $\gamma=7/4$ and $\gamma=4$. These should be viewed as useful benchmarks rather than as two alternative global steady-state profiles. For a single-mass, scattering-dominated population, the zero-flux Bahcall--Wolf solution gives~\cite{1977ApJ...216..883B}
\begin{equation}
n_\bullet(r)\propto r^{-7/4}.
\end{equation}
In a realistic nuclear star cluster,
however, the stellar-mass black holes coexist with a more numerous
population of lighter stars, and the slope of the sBH distribution
depends on which component dominates the relaxation. In the two-component steady-state solution of Linial and Sari, as implemented
for EMRI formation by Rom et al.~\cite{Linial_2022, Rom:2024nso},
the sBHs follow a steep branch
\begin{equation}
n_\bullet(a)\propto a^{-3/2-m_\bullet/(4m_\star)} ,
\end{equation}
which becomes $n_\bullet\propto r^{-4}$ for
$m_\bullet/m_\star\simeq10$, in the region where stars dominate the scattering. At smaller radii, below the transition radius $r_I$, the sBHs themselves dominate relaxation and the sBH profile returns to a
BW-like slope. At still smaller radii, gravitational-wave emission depletes the population, an effect that we model below through the loss-cone/GW-depletion factor.

A useful way to interpret our BW benchmark is therefore in terms of the transition radius $r_I$. In the broken steady-state profile of Ref.~\cite{Rom:2024nso}, where the stars dominate the potential and relaxation outside $r_I$ with an enclosed stellar mass $M_\star(<r_h)= 2 M_\bullet$ fixing their normalisation, the transition radius follows from equating the two species's contributions to the angular-momentum relaxation,
\begin{equation}
\frac{r_I}{r_h}
=
f_\bullet^{4/5}
\left(\frac{m_\star}{m_\bullet}\right)^{-6/5},
\end{equation}
where $f_\bullet$ is the sBH number fraction within the sphere of
influence. Comparing this with the circular-orbit GW-depletion scale
$a_{\rm GW}$ gives, for a Milky-Way-like nucleus,
\begin{equation}
\frac{r_I}{a_{\rm GW}}
\simeq
150
\left(\frac{f_\bullet}{10^{-3}}\right)^{4/5}
\left(\frac{M_\bullet}{4\times10^6M_\odot}\right)^{-1/3}
\left(\frac{m_\bullet/m_\star}{10}\right)^{16/15}
\end{equation}
Thus the intermediate BW interval $a_{\rm GW}<r<r_I$ is not guaranteed to be wide. For $f_\bullet\sim10^{-5}$ it spans only a factor of a few
in radius, while for $f_\bullet\simeq {\rm few}\times10^{-6}$ one has
$R_I\simeq R_{\rm GW}$.

This observation is numerically relevant for the S2/S0-2 normalisations used in this work. For Sgr~A$^*$,
\begin{equation}
a_{\rm GW}\simeq 2\times10^3 R_s\simeq160\,{\rm AU},
\end{equation}
which is very close to our reference radius $r_0=200\,{\rm AU}$. The
BW branch of the Rom et al. profile predicts, for
$m_\bullet=10M_\odot$,
\begin{equation}
n_{\rm BW}^{\rm Rom}(a_0)
\simeq
1.9\times10^{-7}\,{\rm AU}^{-3}.
\end{equation}
Therefore a local normalisation
$n_0\simeq10^{-7}\,{\rm AU}^{-3}$, if interpreted as belonging to the
outer $a^{-4}$ branch, corresponds to
$f_\bullet\simeq {\rm few}\times10^{-6}$ and
$r_I\simeq a_{\rm GW}$. In this low-abundance interpretation, the
steep $r^{-4}$ branch can extend almost down to the GW-depleted region,
so a pure BW extrapolation from the S2/S0-2 scale is conservative for
inner-periapse observables such as EMRBs.

\section{Parabolic-encounter gravitational-wave spectrum}
\label{app:parabolic_spectrum}

In this appendix we summarize the derivation of the gravitational-wave energy
spectrum emitted in a single Newtonian parabolic encounter.  We consider two
masses $M_1$ and $M_2$, total mass $M_{\rm tot}=M_1+M_2$, and periapse
distance $r_p$.  The final result can be written as
\begin{equation}
    \frac{dE}{df}
    =
    \frac{4\pi^2}{5}\frac{G^3}{c^5}
    \frac{M_1^2M_2^2}{r_p^2}\,
    \ell\!\left(\frac{f}{f_c}\right),
    \label{eq:parabolic_spectrum_master}
\end{equation}
where
\begin{equation}
    f_c \equiv \frac{1}{2\pi}
    \left(\frac{G M_{\rm tot}}{r_p^3}\right)^{1/2}.
    \label{eq:parabolic_fc}
\end{equation}
The dimensionless function $\ell$ is obtained as the parabolic limit of the
Peters--Mathews harmonic spectrum.

In fact, for a bound eccentric orbit, the power emitted in the $n$-th harmonic is
\begin{equation}
    P_n(e)
    =
    \frac{32}{5}\frac{G^4}{c^5}
    \frac{M_1^2M_2^2M_{\rm tot}}{a^5}\,
    g(n,e),
    \label{eq:PM_harmonic_power}
\end{equation}
where $a$ is the semimajor axis and $g(n,e)$ is the standard
Peters--Mathews harmonic function~\cite{Peters:1963ux}.  During one radial period
$T=2\pi/\omega_1$, the energy emitted in this harmonic is
\begin{equation}
    E_n = P_n T = \frac{2\pi}{\omega_1}P_n.
\end{equation}
In the parabolic limit $e\to1$ at fixed $r_p=a(1-e)$, the orbital frequency
satisfies
\begin{equation}
    \omega_1^2=\frac{GM_{\rm tot}}{a^3}
    =
    (1-e)^3\omega_c^2,
    \qquad
    \omega_c^2\equiv \frac{GM_{\rm tot}}{r_p^3}.
\end{equation}
The harmonic spacing $\Delta\omega=\omega_1$ therefore tends to zero, and the
discrete spectrum becomes continuous
\begin{equation}
    \frac{dE}{d\omega}\bigg|_{\omega=n\omega_1}
    \simeq
    \frac{E_n}{\omega_1}
    =
    \frac{2\pi}{\omega_1^2}P_n.
\end{equation}
Since $dE/df=2\pi\,dE/d\omega$, one finds
\begin{equation}
    \frac{dE}{df}
    =
    \frac{4\pi^2}{\omega_1^2}P_n
    =
    \frac{128\pi^2}{5}
    \frac{G^3}{c^5}
    \frac{M_1^2M_2^2}{r_p^2}
    (1-e)^2 g(n,e).
    \label{eq:dEdf_before_limit}
\end{equation}
It then remains to take the limit $e\to1,\,n\to\infty$,
with $\tilde f\equiv \frac{n\omega_1}{\omega_c}
    =
    n(1-e)^{3/2}
    =
    \frac{f}{f_c}
$ fixed. We define
\begin{equation}
    \ell(\tilde f)
    \equiv
    \lim_{e\to1}32(1-e)^2g(n,e),
    \label{eq:ell_limit_definition}
\end{equation}
which turns Eq.~\eqref{eq:dEdf_before_limit} into
Eq.~\eqref{eq:parabolic_spectrum_master}.

The required limit is controlled by the large-order, near-turning-point
asymptotics of the Bessel functions appearing in $g(n,e)$.  In particular,
for $n\to\infty$, $e\to1$ with $\tilde f=n(1-e)^{3/2}$ fixed, one has
\begin{equation}
    J_n(ne)
    \simeq
    (1-e)^{1/2}A(\tilde f),
    \qquad
    J_n'(ne)
    \simeq
    (1-e)B(\tilde f).
\end{equation}
Introducing $x \equiv \frac{2^{3/2}}{3}\tilde f$, the two finite functions are
\begin{equation}
    A(\tilde f)
    =
    \frac{1}{\pi}\sqrt{\frac{2}{3}}\,
    K_{1/3}(x),
    \label{eq:A_parabolic}
\end{equation}
and
\begin{equation}
    B(\tilde f)
    =
    \frac{1}{\sqrt{3}\pi}
    \left[
    K_{-2/3}(x)+K_{4/3}(x)
    -
    \frac{1}{\sqrt{2}\,\tilde f}K_{1/3}(x)
    \right].
    \label{eq:B_parabolic}
\end{equation}
Using Bessel recurrence relations, the Peters--Mathews function $g(n,e)$ can
be rewritten in terms of $J_n(ne)$ and $J_n'(ne)$.  Substituting the above
limits gives
\begin{equation}
    \ell(\tilde f)
    =
    \left[
    8\tilde f^2B(\tilde f)-2\tilde f A(\tilde f)
    \right]^2
    +
    \left[
    128\tilde f^4+\frac{4}{3}\tilde f^2
    \right]
    A(\tilde f)^2.
    \label{eq:ell_parabolic}
\end{equation}
Equations~\eqref{eq:parabolic_spectrum_master}--\eqref{eq:ell_parabolic}
are the parabolic-burst spectrum used in the main text.

As a normalization check, integrating over frequency must reproduce the
standard parabolic energy loss,
\begin{equation}
    \Delta E_{\rm par}
    =
    \frac{85\pi}{12\sqrt{2}}
    \frac{G^{7/2}}{c^5}
    \frac{M_1^2M_2^2M_{\rm tot}^{1/2}}{r_p^{7/2}}
    =
    \frac{85\pi}{3\,2^{5/2}}
    \frac{G^3}{c^5}
    \frac{M_1^2M_2^2}{r_p^2}\,
    \omega_c .
    \label{eq:DeltaE_parabolic}
\end{equation}
Indeed,
\begin{equation}
    \Delta E_{\rm par}
    =
    \int_0^\infty df\,\frac{dE}{df}
    =
    \frac{4\pi^2}{5}
    \frac{G^3}{c^5}
    \frac{M_1^2M_2^2}{r_p^2}
    f_c
    \int_0^\infty d\tilde f\,\ell(\tilde f),
\end{equation}
which fixes the overall normalization of $\ell$.  The low-frequency and
high-frequency limits follow directly from the Bessel functions:
\begin{equation}
    \ell(\tilde f)\propto \tilde f^{4/3},
    \qquad
    \tilde f\ll1,
\end{equation}
while for $\tilde f\gg1$ the spectrum is exponentially suppressed.

As demonstrated in Refs.~\cite{Berry:2010gt, Berry:2012im} this approximation accuractely describes the spectral emission in the paramter space of interest, in particular for high eccentricity.

\section{Power-law sensitivity curve for LISA}
\label{app:LISA-PLS}

We model the detector noise as the sum of the instrumental noise and the Galactic white-dwarf confusion foreground, $S_n(f)=S_{\rm inst}(f)+S_{\rm conf}(f)$. For the instrument we adopt the analytic sky-averaged power spectral density of
Ref.~\cite{Robson2019}, constructed from the single-link optical-metrology and test-mass acceleration noises with arm length $L=2.5\times10^{9}\,$m. For the confusion foreground we use their fit for a four-year mission.  The associated characteristic noise strain is $h_c^{\rm n}(f)=\sqrt{f\,S_n(f)}$. Using $h_c^2(f)=f\,S_h(f)$ together with
$\Omega_{\rm GW}(f)=\tfrac{2\pi^2}{3H_0^2}\,f^2\,h_c^2(f)$, the noise
is expressed in the same energy-density units as the signal, $\Omega_n(f)=\tfrac{2\pi^2}{3H_0^2}\,f^3\,S_n(f)$. A stochastic background, however, is a broadband signal accumulated over the mission lifetime, so the relevant figure of merit is not the instantaneous noise floor $\Omega_n$ but the power-law-integrated (PLS) sensitivity~\cite{Thrane2013}. For a power-law
background $\Omega_{\rm GW}(f)=\Omega_\beta\,(f/f_{\rm ref})^\beta$ observed for a time
$T_{\rm obs}$, the optimal signal-to-noise ratio is
$\mathrm{SNR}^2=T_{\rm obs}\!\int\!\mathrm{d}f\,[\Omega_{\rm GW}(f)/\Omega_n(f)]^2$.
For each slope $\beta$ we determine the amplitude $\Omega_\beta$ that yields a fixed
detection threshold $\mathrm{SNR}_{\rm thr}$, and the PLS curve is the envelope of all
such marginally-detectable power laws,
$\Omega_{\rm PLS}(f)=\max_\beta\,\Omega_\beta\,(f/f_{\rm ref})^\beta$. We adopt
$T_{\rm obs}=4\,$yr and $\mathrm{SNR}_{\rm thr}=5$. A
background rising above this curve is detectable at $\mathrm{SNR}\ge5$, which is the
case for our optimistic and fiducial scenarios.

\bibliographystyle{bibi}

\bibliography{biblio}

\providecommand{\href}[2]{#2}\begingroup\raggedright\begin{thebibliography}{10}

\bibitem{Gair:2017ynp}
J.~R. Gair, S.~Babak, A.~Sesana, P.~Amaro-Seoane, E.~Barausse, C.~P.~L. Berry,
  E.~Berti and C.~Sopuerta, \emph{{Prospects for observing extreme-mass-ratio
  inspirals with LISA}},
  \href{https://doi.org/10.1088/1742-6596/840/1/012021}{\emph{J. Phys. Conf.
  Ser.} {\bfseries 840} (2017) 012021}
  [\href{https://arxiv.org/abs/1704.00009}{{\ttfamily 1704.00009}}].

\bibitem{LISA:2024hlh}
{\scshape LISA} Collaboration, M.~Colpi et~al., \emph{{LISA Definition Study
  Report}},  \href{https://arxiv.org/abs/2402.07571}{{\ttfamily 2402.07571}}.

\bibitem{Babak:2017tow}
S.~Babak, J.~Gair, A.~Sesana, E.~Barausse, C.~F. Sopuerta, C.~P.~L. Berry,
  E.~Berti, P.~Amaro-Seoane, A.~Petiteau and A.~Klein, \emph{{Science with the
  space-based interferometer LISA. V: Extreme mass-ratio inspirals}},
  \href{https://doi.org/10.1103/PhysRevD.95.103012}{\emph{Phys. Rev. D}
  {\bfseries 95} (2017) 103012}
  [\href{https://arxiv.org/abs/1703.09722}{{\ttfamily 1703.09722}}].

\bibitem{Amaro-Seoane:2007osp}
P.~Amaro-Seoane, J.~R. Gair, M.~Freitag, M.~Coleman~Miller, I.~Mandel, C.~J.
  Cutler and S.~Babak, \emph{{Astrophysics, detection and science applications
  of intermediate- and extreme mass-ratio inspirals}},
  \href{https://doi.org/10.1088/0264-9381/24/17/R01}{\emph{Class. Quant. Grav.}
  {\bfseries 24} (2007) R113}
  [\href{https://arxiv.org/abs/astro-ph/0703495}{{\ttfamily
  astro-ph/0703495}}].

\bibitem{Bonetti:2020jku}
M.~Bonetti and A.~Sesana, \emph{{Gravitational wave background from extreme
  mass ratio inspirals}},
  \href{https://doi.org/10.1103/PhysRevD.102.103023}{\emph{Phys. Rev. D}
  {\bfseries 102} (2020) 103023}
  [\href{https://arxiv.org/abs/2007.14403}{{\ttfamily 2007.14403}}].

\bibitem{Berry:2010gt}
C.~P.~L. Berry and J.~R. Gair, \emph{{Gravitational wave energy spectrum of a
  parabolic encounter}},
  \href{https://doi.org/10.1103/PhysRevD.82.107501}{\emph{Phys. Rev. D}
  {\bfseries 82} (2010) 107501}
  [\href{https://arxiv.org/abs/1010.3865}{{\ttfamily 1010.3865}}].

\bibitem{Berry:2012im}
C.~P.~L. Berry and J.~R. Gair, \emph{{Observing the Galaxy's massive black hole
  with gravitational wave bursts}},
  \href{https://doi.org/10.1093/mnras/sts360}{\emph{Mon. Not. Roy. Astron.
  Soc.} {\bfseries 429} (2013) 589}
  [\href{https://arxiv.org/abs/1210.2778}{{\ttfamily 1210.2778}}].

\bibitem{Rubbo2006ApJ}
L.~J. {Rubbo}, K.~{Holley-Bockelmann} and L.~S. {Finn}, \emph{{Event Rate for
  Extreme Mass Ratio Burst Signals in the Laser Interferometer Space Antenna
  Band}}, \href{https://doi.org/10.1086/508326}{\emph{Astrophys.J.Lett.}
  {\bfseries 649} (2006) L25}.

\bibitem{Hopman2007}
C.~{Hopman}, M.~{Freitag} and S.~L. {Larson}, \emph{{Gravitational wave bursts
  from the Galactic massive black hole}},
  \href{https://doi.org/10.1111/j.1365-2966.2007.11758.x}{\emph{\mnras}
  {\bfseries 378} (2007) 129}
  [\href{https://arxiv.org/abs/astro-ph/0612337}{{\ttfamily
  astro-ph/0612337}}].

\bibitem{Oliver:2025irg}
D.~J. Oliver, A.~D. Johnson, L.~Janssen, J.~Berrier, K.~Glampedakis and
  D.~Kennefick, \emph{{Gravitational wave peep contributions to background
  signal confusion noise for LISA}},
  \href{https://doi.org/10.1103/kgp7-ymyj}{\emph{Phys. Rev. D} {\bfseries 113}
  (2026) 063001} [\href{https://arxiv.org/abs/2507.19704}{{\ttfamily
  2507.19704}}].

\bibitem{Berry:2013poa}
C.~P.~L. Berry and J.~R. Gair, \emph{{Extreme-mass-ratio-bursts from
  extragalactic sources}},
  \href{https://doi.org/10.1093/mnras/stt990}{\emph{Mon. Not. Roy. Astron.
  Soc.} {\bfseries 433} (2013) 3572}
  [\href{https://arxiv.org/abs/1306.0774}{{\ttfamily 1306.0774}}].

\bibitem{Berry:2013ara}
C.~P.~L. Berry and J.~R. Gair, \emph{{Expectations for extreme-mass-ratio
  bursts from the Galactic Centre}},
  \href{https://doi.org/10.1093/mnras/stt1543}{\emph{Mon. Not. Roy. Astron.
  Soc.} {\bfseries 435} (2013) 3521}
  [\href{https://arxiv.org/abs/1307.7276}{{\ttfamily 1307.7276}}].

\bibitem{Hopman:2005vr}
C.~Hopman and T.~Alexander, \emph{{The Orbital statistics of stellar inspiral
  and relaxation near a massive black hole: Characterizing gravitational wave
  sources}}, \href{https://doi.org/10.1086/431475}{\emph{Astrophys. J.}
  {\bfseries 629} (2005) 362}
  [\href{https://arxiv.org/abs/astro-ph/0503672}{{\ttfamily
  astro-ph/0503672}}].

\bibitem{BarOr2016}
B.~{Bar-Or} and T.~{Alexander}, \emph{{Steady-state Relativistic Stellar
  Dynamics Around a Massive Black hole}},
  \href{https://doi.org/10.3847/0004-637X/820/2/129}{\emph{\apj} {\bfseries
  820} (2016) 129} [\href{https://arxiv.org/abs/1508.01390}{{\ttfamily
  1508.01390}}].

\bibitem{Amaro-Seoane:2012lgq}
P.~Amaro-Seoane, \emph{{Relativistic dynamics and extreme mass ratio
  inspirals}}, \href{https://doi.org/10.1007/s41114-018-0013-8}{\emph{Living
  Rev. Rel.} {\bfseries 21} (2018) 4}
  [\href{https://arxiv.org/abs/1205.5240}{{\ttfamily 1205.5240}}].

\bibitem{Pan:2021ksp}
Z.~Pan and H.~Yang, \emph{{Formation Rate of Extreme Mass Ratio Inspirals in
  Active Galactic Nuclei}},
  \href{https://doi.org/10.1103/PhysRevD.103.103018}{\emph{Phys. Rev. D}
  {\bfseries 103} (2021) 103018}
  [\href{https://arxiv.org/abs/2101.09146}{{\ttfamily 2101.09146}}].

\bibitem{Rom:2024nso}
B.~Rom, I.~Linial, K.~Kaur and R.~Sari, \emph{{Dynamics Around Supermassive
  Black Holes: Extreme-mass-ratio Inspirals as Gravitational-wave Sources}},
  \href{https://doi.org/10.3847/1538-4357/ad8b1d}{\emph{Astrophys. J.}
  {\bfseries 977} (2024) 7} [\href{https://arxiv.org/abs/2406.19443}{{\ttfamily
  2406.19443}}].

\bibitem{Kaur:2024ofj}
K.~Kaur, B.~Rom and R.~Sari, \emph{{Semianalytical Fokker{\textendash}Planck
  Models for Nuclear Star Clusters}},
  \href{https://doi.org/10.3847/1538-4357/ada8a8}{\emph{Astrophys. J.}
  {\bfseries 980} (2025) 150}
  [\href{https://arxiv.org/abs/2406.07627}{{\ttfamily 2406.07627}}].

\bibitem{Alexander:2008tq}
T.~Alexander and C.~Hopman, \emph{{Strong mass segregation around a massive
  black hole}},
  \href{https://doi.org/10.1088/0004-637X/697/2/1861}{\emph{Astrophys. J.}
  {\bfseries 697} (2009) 1861}
  [\href{https://arxiv.org/abs/0808.3150}{{\ttfamily 0808.3150}}].

\bibitem{1977ApJ...216..883B}
J.~N. {Bahcall} and R.~A. {Wolf}, \emph{{The star distribution around a massive
  black hole in a globular cluster. II. Unequal star masses.}},
  \href{https://doi.org/10.1086/155534}{\emph{\apj} {\bfseries 216} (1977)
  883}.

\bibitem{Greene_2020}
J.~E. Greene, J.~Strader and L.~C. Ho, \emph{Intermediate-mass black holes},
  \href{https://doi.org/10.1146/annurev-astro-032620-021835}{\emph{Annual
  Review of Astronomy and Astrophysics} {\bfseries 58} (2020) 257–312}.

\bibitem{Burke:2024wcf}
C.~J. Burke, P.~Natarajan, V.~F. Baldassare and M.~Geha, \emph{{Multiwavelength
  Constraints on the Local Black Hole Occupation Fraction}},
  \href{https://doi.org/10.3847/1538-4357/ad94d9}{\emph{Astrophys. J.}
  {\bfseries 978} (2025) 77}
  [\href{https://arxiv.org/abs/2410.11177}{{\ttfamily 2410.11177}}].

\bibitem{Schodel:2002py}
R.~Schodel et~al., \emph{{A Star in a 15.2 year orbit around the supermassive
  black hole at the center of the Milky Way}},
  \href{https://doi.org/10.1038/nature01121}{\emph{Nature} {\bfseries 419}
  (2002) 694} [\href{https://arxiv.org/abs/astro-ph/0210426}{{\ttfamily
  astro-ph/0210426}}].

\bibitem{Ghez:2008ms}
A.~M. Ghez et~al., \emph{{Measuring Distance and Properties of the Milky Way's
  Central Supermassive Black Hole with Stellar Orbits}},
  \href{https://doi.org/10.1086/592738}{\emph{Astrophys. J.} {\bfseries 689}
  (2008) 1044} [\href{https://arxiv.org/abs/0808.2870}{{\ttfamily 0808.2870}}].

\bibitem{Gillessen_2017}
S.~Gillessen, P.~M. Plewa, F.~Eisenhauer, R.~Sari, I.~Waisberg, M.~Habibi,
  O.~Pfuhl, E.~George, J.~Dexter, S.~v. Fellenberg, T.~Ott and R.~Genzel,
  \emph{An update on monitoring stellar orbits in the galactic center},
  \href{https://doi.org/10.3847/1538-4357/aa5c41}{\emph{The Astrophysical
  Journal} {\bfseries 837} (2017) 30}.

\bibitem{Genzel2010}
R.~{Genzel}, F.~{Eisenhauer} and S.~{Gillessen}, \emph{{The Galactic Center
  massive black hole and nuclear star cluster}},
  \href{https://doi.org/10.1103/RevModPhys.82.3121}{\emph{Reviews of Modern
  Physics} {\bfseries 82} (2010) 3121}
  [\href{https://arxiv.org/abs/1006.0064}{{\ttfamily 1006.0064}}].

\bibitem{GRAVITY2017}
{GRAVITY Collaboration}, R.~{Abuter} et~al., \emph{{First light for GRAVITY:
  Phase referencing optical interferometry for the Very Large Telescope
  Interferometer}},
  \href{https://doi.org/10.1051/0004-6361/201730838}{\emph{A\&A} {\bfseries
  602} (2017) A94} [\href{https://arxiv.org/abs/1705.02345}{{\ttfamily
  1705.02345}}].

\bibitem{GRAVITY:2018ofz}
{\scshape GRAVITY} Collaboration, R.~Abuter et~al., \emph{{Detection of the
  gravitational redshift in the orbit of the star S2 near the Galactic centre
  massive black hole}},
  \href{https://doi.org/10.1051/0004-6361/201833718}{\emph{Astron. Astrophys.}
  {\bfseries 615} (2018) L15}
  [\href{https://arxiv.org/abs/1807.09409}{{\ttfamily 1807.09409}}].

\bibitem{GRAVITY2020}
{GRAVITY Collaboration}, R.~{Abuter} et~al., \emph{{Detection of the
  Schwarzschild precession in the orbit of the star S2 near the Galactic centre
  massive black hole}},
  \href{https://doi.org/10.1051/0004-6361/202037813}{\emph{A\&A} {\bfseries
  636} (2020) L5} [\href{https://arxiv.org/abs/2004.07187}{{\ttfamily
  2004.07187}}].

\bibitem{GRAVITY2024}
{GRAVITY Collaboration} et~al., \emph{{Improving constraints on the extended
  mass distribution in the Galactic Center with stellar orbits}},
  \href{https://doi.org/10.1051/0004-6361/202452274}{\emph{A\&A} {\bfseries
  692} (2024) A242} [\href{https://arxiv.org/abs/2409.12261}{{\ttfamily
  2409.12261}}].

\bibitem{Bordoni:2025mli}
M.~S. Bordoni et~al., \emph{{Impact of a granular mass distribution on the
  orbit of S2 in the Galactic center}},
  \href{https://doi.org/10.1051/0004-6361/202554223}{\emph{Astron. Astrophys.}
  {\bfseries 701} (2025) A89}
  [\href{https://arxiv.org/abs/2507.01510}{{\ttfamily 2507.01510}}].

\bibitem{Preto:2009kd}
M.~Preto and P.~Amaro-Seoane, \emph{{On strong mass segregation around a
  massive black hole: Implications for lower-frequency gravitational-wave
  astrophysics}},
  \href{https://doi.org/10.1088/2041-8205/708/1/L42}{\emph{Astrophys. J. Lett.}
  {\bfseries 708} (2010) L42}
  [\href{https://arxiv.org/abs/0910.3206}{{\ttfamily 0910.3206}}].

\bibitem{Tremaine:2002js}
S.~Tremaine et~al., \emph{{The slope of the black hole mass versus velocity
  dispersion correlation}}, {\emph{Astrophys. J.} {\bfseries 574} (2002) 740}
  [\href{https://arxiv.org/abs/astro-ph/0203468}{{\ttfamily
  astro-ph/0203468}}].

\bibitem{2001MNRAS.322..231K}
P.~{Kroupa}, \emph{{On the variation of the initial mass function}},
  \href{https://doi.org/10.1046/j.1365-8711.2001.04022.x}{\emph{\mnras}
  {\bfseries 322} (2001) 231}
  [\href{https://arxiv.org/abs/astro-ph/0009005}{{\ttfamily
  astro-ph/0009005}}].

\bibitem{Hopman_2006}
C.~Hopman and T.~Alexander, \emph{Resonant relaxation near a massive black
  hole: The stellar distribution and gravitational wave sources},
  \href{https://doi.org/10.1086/504400}{\emph{The Astrophysical Journal}
  {\bfseries 645} (2006) 1152–1163}.

\bibitem{MiraldaEscude2000ApJ}
J.~{Miralda-Escud{\'e}} and A.~{Gould}, \emph{{A Cluster of Black Holes at the
  Galactic Center}}, \href{https://doi.org/10.1086/317837}{\emph{\apj}
  {\bfseries 545} (2000) 847}
  [\href{https://arxiv.org/abs/astro-ph/0003269}{{\ttfamily
  astro-ph/0003269}}].

\bibitem{2006ApJ...649...91F}
M.~{Freitag}, P.~{Amaro-Seoane} and V.~{Kalogera}, \emph{{Stellar Remnants in
  Galactic Nuclei: Mass Segregation}},
  \href{https://doi.org/10.1086/506193}{\emph{\apj} {\bfseries 649} (2006) 91}
  [\href{https://arxiv.org/abs/astro-ph/0603280}{{\ttfamily
  astro-ph/0603280}}].

\bibitem{Hopman:2009}
C.~Hopman, \emph{{Extreme mass ratio inspiral rates: dependence on the massive
  black hole mass}}, {\emph{Astrophys. J.} {\bfseries 700} (2009) 1933}
  [\href{https://arxiv.org/abs/0901.1667}{{\ttfamily 0901.1667}}].

\bibitem{Barausse:2012fy}
E.~Barausse, \emph{{The evolution of massive black holes and their spins in
  their galactic hosts}},
  \href{https://doi.org/10.1111/j.1365-2966.2012.21057.x}{\emph{Mon. Not. Roy.
  Astron. Soc.} {\bfseries 423} (2012) 2533}
  [\href{https://arxiv.org/abs/1201.5888}{{\ttfamily 1201.5888}}].

\bibitem{Pozzoli:2023kxy}
F.~Pozzoli, S.~Babak, A.~Sesana, M.~Bonetti and N.~Karnesis, \emph{{Computation
  of stochastic background from extreme-mass-ratio inspiral populations for
  LISA}}, \href{https://doi.org/10.1103/PhysRevD.108.103039}{\emph{Phys. Rev.
  D} {\bfseries 108} (2023) 103039}
  [\href{https://arxiv.org/abs/2302.07043}{{\ttfamily 2302.07043}}]. [Erratum:
  Phys.Rev.D 110, 049903 (2024)].

\bibitem{Peters:1963ux}
P.~C. Peters and J.~Mathews, \emph{{Gravitational radiation from point masses
  in a Keplerian orbit}},
  \href{https://doi.org/10.1103/PhysRev.131.435}{\emph{Phys. Rev.} {\bfseries
  131} (1963) 435}.

\bibitem{1987degc.book.....S}
L.~{Spitzer}, \emph{{Dynamical evolution of globular clusters}}. 1987.

\bibitem{binney1987}
J.~Binney and S.~Tremaine, \emph{Galactic Dynamics}. Princeton Series in
  Astrophysics, 1987.

\bibitem{Qunbar:2023vys}
I.~Qunbar and N.~C. Stone, \emph{{Enhanced Extreme Mass Ratio Inspiral Rates
  and Intermediate Mass Black Holes}},
  \href{https://doi.org/10.1103/PhysRevLett.133.141401}{\emph{Phys. Rev. Lett.}
  {\bfseries 133} (2024) 141401}
  [\href{https://arxiv.org/abs/2304.13062}{{\ttfamily 2304.13062}}].

\bibitem{Mancieri:2024sfy}
D.~Mancieri, L.~Broggi, M.~Bonetti and A.~Sesana, \emph{{Hanging on the cliff:
  Extreme mass ratio inspiral formation with local two-body relaxation and
  post-Newtonian dynamics}},
  \href{https://doi.org/10.1051/0004-6361/202452306}{\emph{Astron. Astrophys.}
  {\bfseries 694} (2025) A272}
  [\href{https://arxiv.org/abs/2409.09122}{{\ttfamily 2409.09122}}].

\bibitem{Phinney:2001di}
E.~S. Phinney, \emph{{A Practical theorem on gravitational wave backgrounds}},
  \href{https://arxiv.org/abs/astro-ph/0108028}{{\ttfamily astro-ph/0108028}}.

\bibitem{Toonen:2009qi}
S.~Toonen, C.~Hopman and M.~Freitag, \emph{{The gravitational wave background
  from star-massive black hole fly-bys}},
  \href{https://doi.org/10.1111/j.1365-2966.2009.15204.x}{\emph{Mon. Not. Roy.
  Astron. Soc.} {\bfseries 398} (2009) 1228}
  [\href{https://arxiv.org/abs/0902.3253}{{\ttfamily 0902.3253}}].

\bibitem{Oliver:2023xan}
D.~J. Oliver, A.~D. Johnson, J.~Berrier, K.~Glampedakis and D.~Kennefick,
  \emph{{Gravitational wave peeps from EMRIs and their implication for LISA
  signal confusion noise}},
  \href{https://doi.org/10.1088/1361-6382/ad40f2}{\emph{Class. Quant. Grav.}
  {\bfseries 41} (2024) 115004}
  [\href{https://arxiv.org/abs/2305.05793}{{\ttfamily 2305.05793}}].

\bibitem{1977ApJ...211..244L}
A.~P. {Lightman} and S.~L. {Shapiro}, \emph{{The distribution and consumption
  rate of stars around a massive, collapsed object.}},
  \href{https://doi.org/10.1086/154925}{\emph{\apj} {\bfseries 211} (1977)
  244}.

\bibitem{Cohn1978ApJ}
H.~{Cohn} and R.~M. {Kulsrud}, \emph{{The stellar distribution around a black
  hole: numerical integration of the Fokker-Planck equation.}},
  \href{https://doi.org/10.1086/156685}{\emph{\apj} {\bfseries 226} (1978)
  1087}.

\bibitem{Hopman:2006fc}
C.~Hopman, M.~Freitag and S.~L. Larson, \emph{{Gravitational wave bursts from
  the Galactic massive black hole}},
  \href{https://doi.org/10.1111/j.1365-2966.2007.11758.x}{\emph{Mon. Not. Roy.
  Astron. Soc.} {\bfseries 378} (2007) 129}
  [\href{https://arxiv.org/abs/astro-ph/0612337}{{\ttfamily
  astro-ph/0612337}}].

\bibitem{Thrane2013}
E.~Thrane and J.~D. Romano, \emph{{Sensitivity curves for searches for
  gravitational-wave backgrounds}},
  \href{https://doi.org/10.1103/PhysRevD.88.124032}{\emph{Phys. Rev. D}
  {\bfseries 88} (2013) 124032}
  [\href{https://arxiv.org/abs/1310.5300}{{\ttfamily 1310.5300}}].

\bibitem{Robson2019}
T.~Robson, N.~J. Cornish and C.~Liu, \emph{{The construction and use of LISA
  sensitivity curves}},
  \href{https://doi.org/10.1088/1361-6382/ab1101}{\emph{Class. Quantum Grav.}
  {\bfseries 36} (2019) 105011}
  [\href{https://arxiv.org/abs/1803.01944}{{\ttfamily 1803.01944}}].

\bibitem{Aller:2002rp}
M.~C. Aller and D.~Richstone, \emph{{The Cosmic density of massive black holes
  from galaxy velocity dispersions}},
  \href{https://doi.org/10.1086/344484}{\emph{Astron. J.} {\bfseries 124}
  (2002) 3035} [\href{https://arxiv.org/abs/astro-ph/0210573}{{\ttfamily
  astro-ph/0210573}}].

\bibitem{OLeary:2008myb}
R.~M. O'Leary, B.~Kocsis and A.~Loeb, \emph{{Gravitational waves from
  scattering of stellar-mass black holes in galactic nuclei}},
  \href{https://doi.org/10.1111/j.1365-2966.2009.14653.x}{\emph{Mon. Not. Roy.
  Astron. Soc.} {\bfseries 395} (2009) 2127}
  [\href{https://arxiv.org/abs/0807.2638}{{\ttfamily 0807.2638}}].

\bibitem{Sesana:2019vho}
A.~Sesana et~al., \emph{{Unveiling the gravitational universe at $\mu$-Hz
  frequencies}}, \href{https://doi.org/10.1007/s10686-021-09709-9}{\emph{Exper.
  Astron.} {\bfseries 51} (2021) 1333}
  [\href{https://arxiv.org/abs/1908.11391}{{\ttfamily 1908.11391}}].

\bibitem{Ni:2012eh}
W.-T. Ni, \emph{{ASTROD-GW: Overview and Progress}},
  \href{https://doi.org/10.1142/S0218271813410046}{\emph{Int. J. Mod. Phys. D}
  {\bfseries 22} (2013) 1341004}
  [\href{https://arxiv.org/abs/1212.2816}{{\ttfamily 1212.2816}}].

\bibitem{AmaroSeoane:2011zz}
P.~Amaro-Seoane and M.~Preto, \emph{{The impact of realistic models of mass
  segregation on the event rate of extreme-mass ratio inspirals and cusp
  re-growth}}, {\emph{Class. Quant. Grav.} {\bfseries 28} (2011) 094017}
  [\href{https://arxiv.org/abs/1010.5781}{{\ttfamily 1010.5781}}].

\bibitem{Emami:2019uty}
R.~Emami and A.~Loeb, \emph{{Detectability of gravitational waves from a
  population of inspiralling black holes in Milky Way-mass galaxies}},
  \href{https://doi.org/10.1093/mnras/stab290}{\emph{Mon. Not. Roy. Astron.
  Soc.} {\bfseries 502} (2021) 3932}
  [\href{https://arxiv.org/abs/1903.02579}{{\ttfamily 1903.02579}}].

\bibitem{Emami:2019mzi}
R.~Emami and A.~Loeb, \emph{{Observational signatures of the black hole mass
  distribution in the galactic center}},
  \href{https://doi.org/10.1088/1475-7516/2020/02/021}{\emph{JCAP} {\bfseries
  02} (2020) 021} [\href{https://arxiv.org/abs/1903.02578}{{\ttfamily
  1903.02578}}].

\bibitem{Maggiore:2007ulw}
M.~Maggiore, \emph{{Gravitational Waves. Vol. 1: Theory and Experiments}}.
  Oxford University Press, 2007,
  \href{https://doi.org/10.1093/acprof:oso/9780198570745.001.0001}{10.1093/acprof:oso/9780198570745.001.0001}.

\bibitem{Barack:2004wc}
L.~Barack and C.~Cutler, \emph{{Confusion noise from LISA capture sources}},
  \href{https://doi.org/10.1103/PhysRevD.70.122002}{\emph{Phys. Rev. D}
  {\bfseries 70} (2004) 122002}
  [\href{https://arxiv.org/abs/gr-qc/0409010}{{\ttfamily gr-qc/0409010}}].

\bibitem{Dayem:2026ktt}
K.~A.~E. Dayem et~al., \emph{{Discovery of a star sensitive to the spin of Sgr
  A*}},  \href{https://arxiv.org/abs/2607.12664}{{\ttfamily 2607.12664}}.

\bibitem{EPTA:2023xxk}
{\scshape EPTA, InPTA} Collaboration, J.~Antoniadis et~al., \emph{{The second
  data release from the European Pulsar Timing Array - IV. Implications for
  massive black holes, dark matter, and the early Universe}},
  \href{https://doi.org/10.1051/0004-6361/202347433}{\emph{Astron. Astrophys.}
  {\bfseries 685} (2024) A94}
  [\href{https://arxiv.org/abs/2306.16227}{{\ttfamily 2306.16227}}].

\bibitem{NANOGrav:2023hfp}
{\scshape NANOGrav} Collaboration, G.~Agazie et~al., \emph{{The NANOGrav 15 yr
  Data Set: Constraints on Supermassive Black Hole Binaries from the
  Gravitational-wave Background}},
  \href{https://doi.org/10.3847/2041-8213/ace18b}{\emph{Astrophys. J. Lett.}
  {\bfseries 952} (2023) L37}
  [\href{https://arxiv.org/abs/2306.16220}{{\ttfamily 2306.16220}}].

\bibitem{Sato-Polito:2023gym}
G.~Sato-Polito, M.~Zaldarriaga and E.~Quataert, \emph{{Where are the
  supermassive black holes measured by PTAs?}},
  \href{https://doi.org/10.1103/PhysRevD.110.063020}{\emph{Phys. Rev. D}
  {\bfseries 110} (2024) 063020}
  [\href{https://arxiv.org/abs/2312.06756}{{\ttfamily 2312.06756}}].

\bibitem{Goncharov:2024htb}
B.~Goncharov et~al., \emph{{Reading signatures of supermassive binary black
  holes in pulsar timing array observations}},
  \href{https://doi.org/10.1038/s41467-025-65450-3}{\emph{Nature Commun.}
  {\bfseries 16} (2025) 9692}
  [\href{https://arxiv.org/abs/2409.03627}{{\ttfamily 2409.03627}}].

\bibitem{Sesana:2025udx}
A.~Sesana and D.~G. Figueroa, \emph{{Nanohertz Gravitational Waves}},
  \href{https://arxiv.org/abs/2512.18822}{{\ttfamily 2512.18822}}.

\bibitem{Linial_2022}
I.~Linial and R.~Sari, \emph{Stellar distributions around a supermassive black
  hole: Strong-segregation regime revisited},
  \href{https://doi.org/10.3847/1538-4357/ac9bfd}{\emph{The Astrophysical
  Journal} {\bfseries 940} (2022) 101}.

\end{thebibliography}\endgroup

\end{document}